\documentclass[onecolumn]{ametsocV6.1}
\usepackage{soul}
\usepackage[final]{pdfpages}

\title{Predicting Tropical Cyclone Track Forecast Errors using a Probabilistic Neural Network}

\authors{
    M.A. Fernandez,\aff{a} \correspondingauthor{M.A. Fernandez, mafern@colostate.edu}
    Elizabeth A. Barnes,\aff{a}
    Randal J. Barnes,\aff{b}
    Mark DeMaria,\aff{c}
    Marie McGraw,\aff{c}
    Galina Chirokova,\aff{c}
    and Lixin Lu\aff{c}
    }

\affiliation{
    \aff{a}{Department of Atmospheric Science, Colorado State University, Fort Collins, CO, USA}\\
    \aff{b}{Department of Civil, Environmental, and Geo-Engineering, University of Minnesota, Minneapolis, MN, USA}\\
    \aff{c}{Cooperative Institute for Research in the Atmosphere, Colorado State University, Fort Collins, CO, USA}
    }

\abstract{
    A new method for estimating tropical cyclone track uncertainty is presented and tested.
    This method uses a neural network to predict a bivariate normal distribution, which serves as an estimate for track uncertainty.
    We train the network and make predictions on forecasts from the National Hurricane Center (NHC), which currently uses static error distributions based on forecasts from the past five years for most applications.
    The neural network-based method produces uncertainty estimates that are dynamic and probabilistic.
    Further, the neural network-based method allows for probabilistic statements about tropical cyclone trajectories, including landfall probability, which we highlight.
    We show that our predictions are well calibrated using multiple metrics, that our method produces better uncertainty estimates than current NHC approaches, and that our method achieves similar performance to the Global Ensemble Forecast System.
    Once trained, the computational cost of predictions using this method is negligible, making it a strong candidate to improve the NHC's operational estimations of tropical cyclone track uncertainty.
    }

\begin{document}
\maketitle

\statement
Tropical cyclones affect millions of people across the planet, and accurate uncertainty estimates for their trajectories are vital for informing risk, evacuations, and mitigation planning.
For most applications, the National Hurricane Center currently quantifies uncertainty using a historical-based estimate that remains static for the entire season.
We propose a method that uses machine learning to dynamically estimate track uncertainty using inputs that are specific to the storm being forecast.
Our method produces a probability distribution, specifically a bivariate normal, which presents decision-makers and researchers with a more informative assessment of tropical cyclone track uncertainty.
We demonstrate that our method has many appealing properties, including the ability to produce landfall probabilities and outperform currently used National Hurricane Center methods.

\section[Introduction]{Introduction}\label{sec:introduction}

Tropical cylcones (TCs) expose populations and assets to risk around the world, with negative effects on population well-being \citep{Berlemann2021}.
Forecasting centers have improved TC track accuracy through better modeling and techniques \citep{Heming2019}, though TC track uncertainty has not undergone similar improvements \citep{Dunion2023}.
Uncertainty quantification is particulary important for TC forecasting, informing risk assessment and disaster planning (including mitigation and evacuations) on times scales from hours to days, as well as policy decisions when aggregated over entire TC seasons. 
Here, we focus on estimating TC track uncertainty directly, rather than making predictions of TC tracks with uncertainty as a byproduct.

The National Weather Service (NWS) National Hurricane Center (NHC) has long recognized the need to provide uncertainty information with their deterministic TC forecasts.
The first NHC track uncertainty product was the Strike Probabilities, which became operational in 1983 \citep{Sheets1985}.
The Strike Probabilities only considered track uncertainty and were replaced by wind speed probabilities (WSP) in 2006, which take into account the uncertainty in the track, intensity and wind structure forecasts \citep{Demaria2009}.
The NHC also provides the graphical ``cone of uncertainty,'' which shows the area enclosed within the 67th percentile of the NHC's historical track error distributions (for a given year, the historical is the previous 5 years).
In addition, the NHC’s storm surge watches and warnings implemented in 2017 are based on a probabilistic storm surge model \citep[P-surge,][]{Penny2023}, which uses an ensemble of statistically generated track and intensity forecasts to drive a simplified surge model.

Although uncertainty information is included in many NHC products, the underlying probabilities are determined almost entirely from historical error distributions and include little information about the specific forecast situation.
For example, the variability in the wind forcing for the P-surge model \citep{Penny2023} and the track and intensity uncertainty for the WSP model are based on historical forecast errors from the NHC, the Central Pacific Hurricane Center (CPHC), or the Joint Typhoon Warning Center (JTWC) from the past 5 years.
A method to add situational-dependence to the track uncertainy was added to the WSP model in 2011 by stratifying the NHC track errors by the Goerss Predicted Consensus Error \citep[GPCE,][]{Goerss2007}.
GPCE uses linear regression to predict the track error of a consensus model based on the spread of the models in the consensus and the TC intensity.
However, the wind structure variability in the WSP model is still determined from historical error distributions, and the GPCE input only has a small impact on the track error distributions \citep{DeMaria2013}.

Another weakness of current operational track forecast uncertainty estimates is that the error estimates are circular.
For example, the NHC cone of uncertainty uses static radii for each forecast basin (Atlantic, Eastern Pacific, Central Pacific) and forecast time so that for each point along the forecast track the error estimate is a circle.
The GPCE product estimates the expected error of the consensus forecast and then scales that to a radius that includes the forecast track $\sim68\%$ of the time, similar to that used in the cone.
\cite{Hansen2011} developed a generalized version of GPCE called GPCE-AX (GPCE along–across) that includes separate regression equations for the along- and across-track errors so that the uncertainy areas are not circular.
However, GPCE-AX is not used operationally by the NHC or CPHC in any of their public-facing forecast uncertainty products.

A probabilistic method that has been explored previously is the use of ensemble forecasts for estimating TC track and track uncertainty \citep{Dupont2011, Bonnardo2019, Yasuhiro2020, Zhang2017, Dunion2023, Wilks2009}.
Ensemble systems, such as those based on the Global Forecasting System (GFS) and European Centre for Medium-Range Weather Forecasts (ECMWF) global models, support forecasters in understanding possible track scenarios when making their deterministic track forecasts.
An example of such an ensemble system is the Global Ensemble Forecast System \citep[GEFS,][]{Zhou2017, Guan2022}, which is based on the GFS.
While ensemble systems provide useful information, the public-facing probabilistic products from operational centers need to be consistent with their deterministic forecasts.
For example, if the NHC track forecast shows a landfall in Miami, but all or most of the ensemble members are north of that position, the contradiction between the products could cause considerable confusion.
Therefore, corrections to ensemble systems are needed if they are used for public-facing uncertainty products.

Inclusion of situationally-dependent forecast uncertainty in NWS products remains a high priority\footnote{\url{https://www.weather.gov/media/wrn/NWS-2023-Strategic-Plan.pdf}}.
An emerging method for estimating uncertainty is through the use of machine learning methods \citep{Haynes2023, BarnesBarnes2021, Foster2021, GuillauminZanna2021, GordonBarnes2022}.
Recent work by \cite{barnes_barnes_demaria_2023} used an artificial neural network to predict the parameters of a probability distribution as a means of quantifying uncertainty \citep{NixWeigend1994, NixWeigend1995} for TC intensity forecasting, with applications to rapid intensification prediction.
Here, we ask whether we can make meaningful predictions of TC track uncertainty for specific TCs in a well-calibrated probabilistic framework.

To answer that question, we use a similar framework to \cite{barnes_barnes_demaria_2023}.
We task a neural network with predicting the parameters of a distribution (in this case a bivariate normal distribution) that estimates TC track latitude and longitude uncertainty in kilometers.
Specifically, our framework is designed for use in NHC operations, i.e., we have trained and tested our method on official NHC forecasts, so that the uncertainty products will maintain consistency with the NHC official forecast.

There are many benefits to quantifying uncertainty using the approach detailed in this work.
Like the historical-based measures of uncertainty, our predictions are data-driven: in this case, we use a neural network.
Unlike the current operational methods, our bivariate normal predictions are based on forecast-specific inputs, including environmental variables and dynamical model outputs, and can vary through the correlation and two variance parameters.
The use of a defined distribution means our method does not require running expensive ensembles, or calculating statistics from a limited population of ensemble members, but can use output from those systems as input to the network.
Because of the probabilistic approach and forecast-specific inputs, the predictions returned by this network are a plausible alternative to the historically-derived track uncertainty estimates used in NHC and CPHC operations.

\section{Prediction Framework}\label{sec:methods}

Our goal is to make well-calibrated probabilistic predictions of TC track uncertainty using forecast-specific inputs.
We accompish this task using a neural network that predicts the five parameters of a bivariate normal distribution, which serves as our estimated TC track uncertainty.
The bivariate normal distribution was also used for the original Strike Probability product.
Other choices for the distribution were explored; however, the bivariate normal was effective and simplicity won out over other potential choices.
The dataset we use includes forecast-specific inputs and true errors from NHC and CPHC forecasts covering seasons from 2013 through 2023, a period chosen to balance model availability and quality with sample size.
In the remainder of the discussion, the term NHC is assumed to include the NHC and CPHC forecasts.

\subsection{Neural Network}\label{sec:neural-networks}

\begin{figure}[t]
    \begin{center}
        \noindent\includegraphics[width=\textwidth]{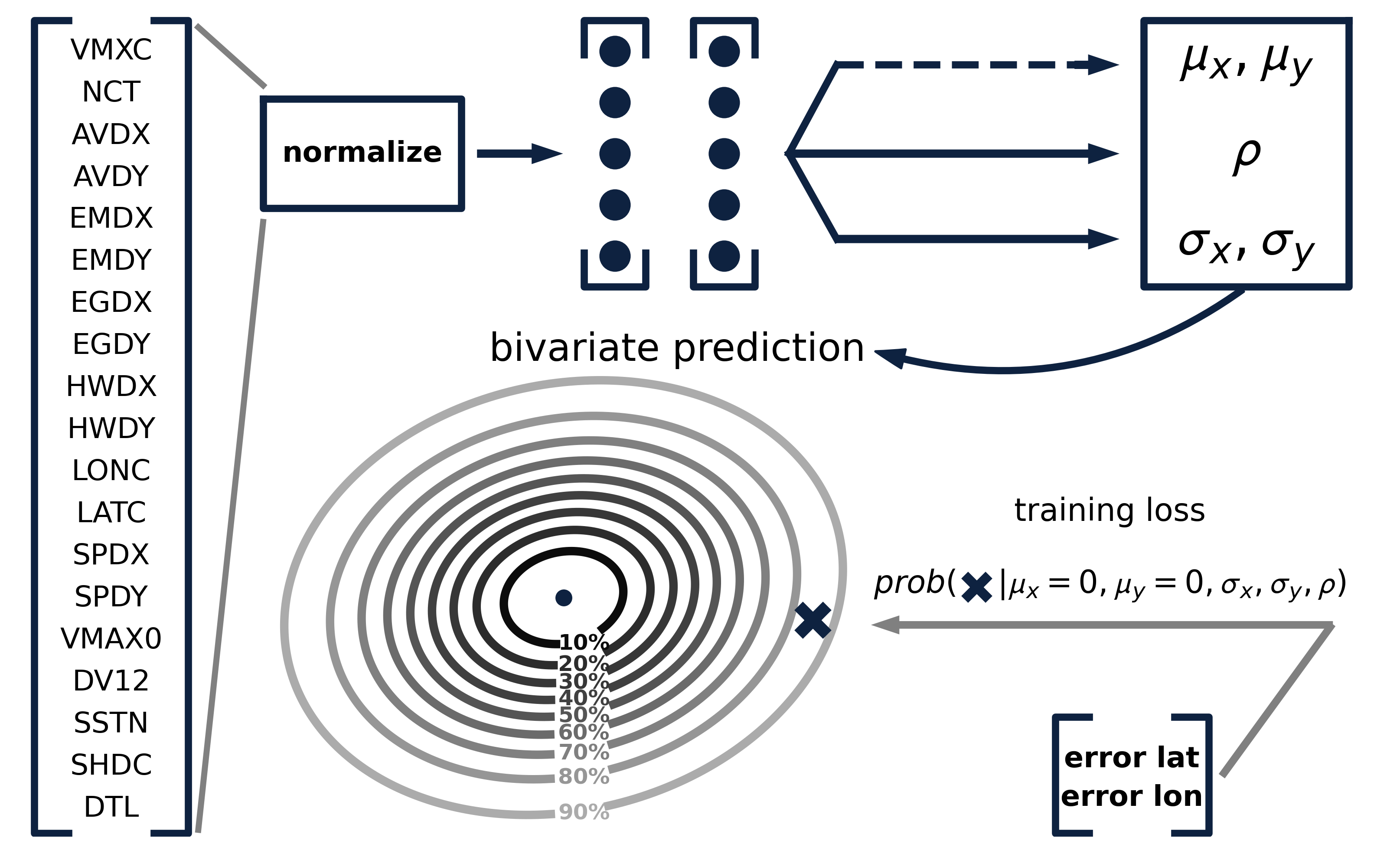}
        \caption{
            Schematic showing the network architecture and format of model predictions.
            In many of the following figures, these predictions are used to construct a two-dimensional cumulative distribution function, defined by the Mahalanobis distance \citep{Mahalanobis}.
            Each labeled ellipse encloses the integrated probability out to that distance.
            Larger percentiles enclose more of the probability, thus, confidence that the truth falls within a percentile increases with percentile.
            Inputs are described in Table~\ref{tab:inputs}.
                }\label{fig:prediction_schematic}
    \end{center}
\end{figure}

The full prediction framework is shown schematically in Figure \ref{fig:prediction_schematic}.
A set of forecast-specific model-based and environmental variables is first normalized by removing the mean and dividing by the standard deviation before being passed to the dense layers of the network.
A detailed list and description of these inputs are shown in Table \ref{tab:inputs}.
There are two fully connected dense layers with five nodes each, both of which use the rectified linear (ReLu) activation function.

The outputs, which are the five parameters of a bivariate normal distribution in two dimensions (subscript $x$ and $y$ represent longitude and latitude), are then each handled separately.
The standard deviations ($\sigma_x, \sigma_y$) are passed through a soft plus layer, which has the form $\ln(1+e^{\alpha})$, and shifts the range to $[0, \infty]$.
The correlation ($\rho$) is passed through a hyperbolic tangent layer (tanh), which shifts the range to $[-1, 1]$.
There is no range restriction on the means ($\mu_x, \mu_y$).
Non-zero $\mu_x$ and $\mu_y$ act as corrections to the NHC forecast for TC latitude and longitude.
Our priority is to produce meaningful estimates for track uncertainty, so in all of our results we freeze $\mu_x$ and $\mu_y$ to zero.
Testing without this restriction showed that the predicted $\mu_x$ and $\mu_y$ are small.
This is not surprising since the track bias of the NHC forecasts is generally much smaller than the mean track error for large samples.
All of the parameters are then rescaled to return to their original units immediately prior to output.

The network trains by minimizing the loss defined by the negative log probability of the NHC forecast error (i.e., the truth, as shown in the lower right bracketed list in Figure \ref{fig:prediction_schematic}), given the predicted bivariate normal distribution.
Note that in Figure \ref{fig:prediction_schematic} the example bivariate normal shown is the cumulative distribution (CDF), rather than the probability density function (PDF) from which the loss is calculated.
The loss function penalizes narrow predictions when the true forecast error is large (i.e., the truth lies outside the bulk of the distribution), and penalizes broad predictions when the true forecast error is small (i.e., the entire distribution is relatively flat).
Early stopping is used for the training, with a patience of 250 epochs.
The batch size is 64, with a learning rate of 0.0001.

The data is separated into the Atlantic and Eastern/Central Pacific basins.
There are less than $2000$ Central Pacific samples in the dataset, which is too small to reasonably train a separate network on.
We trained networks without including the Central Pacific samples and found only marginal changes to the Eastern Pacific estimates, thus we combined these basins.
The data is further separated into lead times every 12 hours up to five days.
A separate network is trained for each basin and lead time combination.
We tested the effectiveness of predicting all lead times using a single network, where lead time was used as an input feature.
Predictions made using this set-up generally evolved more smoothly over lead time, but tended toward more circular predictions (i.e., the correlation parameter $\rho$ was consistently near zero) and did not noticeably improve or degrade the predictions overall (not shown).

For each network, the data is split into training, validation, and testing sets.
The testing set, which is all samples from a given year, is split off first.
The validation set is 200 randomly selected samples from the remaining data, and the training set is the rest of the samples.
For the results shown in this work, we use leave-one-year-out, which iterates through all potential years for the testing set.
Thus the total number of trained networks is 220 (two basins, 10 lead times, 11 years).
In this way, we make predictions for all forecasts over the entire data set without ever using the testing samples for training.

While our predictions are flexible and dynamic, the method of producing those predictions (neural networks) is opaque.
In Supplementary section S4 we use the exlpainable artificial intelligence (XAI) method SHapley Additive exPlanations \citep[SHAP;][]{SHAP} to explore feature relevance, i.e., how our network arrived at its predictions.

\subsection{Dataset}\label{sec:data}

The data set used for labels (truth) and inputs is listed in Tables \ref{tab:labels} and \ref{tab:inputs}.
Each of these variables is recorded for official forecasts made by the NHC during the 2013-2023 seasons, with potential lead times from 12 to 120 hours.
The NHC currently does not make forecasts for 84 and 108 hour lead times and did not make 60 hour lead time forecasts until 2019.
NHC track points at these times were obtained by linear interpolation if an NHC forecast was available before and after each of those times.
The forecast time in all cases is based on synoptic time (i.e., 0000, 0600, 1200, 1800 hours UTC).
The data set includes 186 TCs in the Atlantic and 217 TCs in the Eastern/Central Pacific, for a total of over 40,000 forecasts.
This is an updated version (now including the 2022 and 2023 seasons) of the raw data set used in \cite{barnes_barnes_demaria_2023} to predict TC intensity.

Labels (i.e., the truth) are derived from the best track verification, a post-storm analysis that includes the TC track among other TC characteristics \citep{LandseaFranklin2013}.
The labels are the distance, in kilometers, between the best track latitude and longitude and the relevant forecasted latitude and longitude, i.e., the error between forecast and true TC location.

\begin{table}[t]
    \caption{
        Short name and description of the label variables (i.e., the truth) used in training our networks.
    }\label{tab:labels}
    \begin{center}
        \begin{tabular}{p{4cm}p{9cm}}
            \hline\hline
            label variable & \hfill description \\
            \hline
            OFDX  [km] & \hfill distance east of the best track position from the NHC official forecast \\
            OFDY  [km] & \hfill distance north of the best track position from the NHC official forecast \\
            \hline
        \end{tabular}
    \end{center}
\end{table}

Inputs include both dynamical model forecasts and TC predictors from statistical models.
The TC predictors were included because the performance of the dynamical models have some dependence on these.
For example, one of the most significant predictors of track error in GPCE is the TC intensity.

The dynamical models were chosen based on their track forecast skill and the availability of a long data record for training.
Based on these criteria, one regional hurricane model and three global models were included as follows: HWRF \citep[Hurricane Weather Research and Forecast Model,][]{Tallapragada2016}, UKMet global model \citep{Bush2023}, GFS \citep{Zhou2019}, and ECMWF \citep{Magnusson2021}. 
Outputs from these models are not available until after the NHC official forecast is issued, thus, an interpolated version (based on the previous forecast cycle) is used as input.
The interpolated models are referred to as ``early'' models. 
We use the early models to be consistent with what is available to NHC forecasters at advisory time \citep{Cangialosi2023} and so the uncertainty estimates can be determined shortly after the advisories are issued.

The average of the four early track model forecasts is called the ``consensus'' and can be calculated as long as at least two of the four input models are available at a given forecast time.
The neural network inputs from the early models are the deviations from the consensus forecast (AVDX through HWDY in Table \ref{tab:inputs}).
For missing models, the track forecast is replaced by the consensus of the available models, so the deviations are zero for that model.
The number of models (NCT) is also included as a predictor because the track errors might be larger when some of the skillful models are not available.
Most of the dataset has all four models available ($72\%$), with a small percentage having fewer than two models ($2.5\%$).

The TC predictors include three basic TC parameters (latitude and longitude of the TC center and the maximum wind).
The latitude and longitude are from the consensus forecast (LATC and LONC), and the maximum wind (VMXC) is from a consensus of four skillful early intensity models comprised of the GFS, HWRF and two statistical dynamical intensity models \citep{barnes_barnes_demaria_2023}.

Seven additional TC predictors are obtained from the statistical-dynamical D-SHIPS \citep[Decay-Statistical Hurricane Intensity Prediction Scheme,][]{DeMaria2022}.
These are comprised of the zero hour maximum wind (VMAX0, sustained one-minute average estimate at synoptic time), the change in maximum wind over the 12 hour period ending at the start of the forecast (DV12), the eastward and northward components of the TC translational velocity (SPDX, SPDY), the distance of the TC center from major landmasses (DTL), the sea surface temperature (SSTN), and the 850 to 200 hPa wind shear averaged from 0-500 kilometers (SHDC).
The last six of the above predictors require a track forecast, which is obtained from an interpolated (early) version of the NHC official forecast from the previous cycle in the D-SHIPS model, which is often run prior to the official TC genesis declaration.

\begin{table}[t]
    \caption{
        Input variables used in training our networks.
        }\label{tab:inputs}
    \begin{center}
        \begin{tabular}{p{4cm}p{9cm}}
            \hline\hline
            input variable & \hfill description \\
            \hline
            VMXC  [kt]          & \hfill max wind of the consensus forecast \\
            NCT   [\#]          & \hfill number of models included in the consensus forecast \\
            AVDX  [km]          & \hfill distance east of the early GFS forecast from the consensus forecast \\
            AVDY  [km]          & \hfill distance north of the early GFS forecast from the consensus forecast \\
            EMDX  [km]          & \hfill distance east of the early ECMWF forecast from the consensus forecast \\
            EMDY  [km]          & \hfill distance north of the early ECMWF forecast from the consensus forecast \\
            EGDX  [km]          & \hfill distance east of the early UKMet forecast from the consensus forecast \\
            EGDY  [km]          & \hfill distance north of the early UKMet forecast from the consensus forecast \\
            HWDX  [km]          & \hfill distance east of the early HWRF forecast from the consensus forecast \\
            HWDY  [km]          & \hfill distance north of the early HWRF forecast from the consensus forecast \\
            LONC  [deg E]       & \hfill longitude of the consensus forecast \\
            LATC  [deg N]       & \hfill latitude of the consensus forecast \\
            SPDX  [kt]          & \hfill average eastward speed from DSHP in the 24 hours preceding the forecast \\
            SPDY  [kt]          & \hfill same as SPDX for the northward speed of the TC \\
            VMAX0 [kt]          & \hfill max wind at the start of the forecast \\
            DV12  [kt]          & \hfill intensity change in the 12 hours preceding the forecast \\
            SSTN  [$^{\circ}$C] & \hfill average SST in the 24 hours preceding the forecast \\
            SHDC  [kt]          & \hfill average 850-200 hPa vertical shear in the 24 hours preceding the forecast \\
            DTL   [km]          & \hfill distance to the nearest major landmass at forecast time \\
            \hline
        \end{tabular}
    \end{center}
\end{table}

Analyses and figures presented in this work use our estimates of the uncertainty of the NHC official forecast, which could be used as input for other hazard products such as NHC's wind speed probability or P-surge models.
However, predictions can also be made with respect to the consensus forecast, which would be available before the NHC forecast is issued due to the use of early model input.
The consensus uncertainty could be used as guidance by NHC forecasters for their official forecasts and products such as the Tropical Cyclone Discussion, which sometimes include qualitative descriptions of forecast confidence\footnote{\url{https://www.nhc.noaa.gov/aboutnhcprod.shtml}}.

\subsection{Model Calibration}\label{sec:calibration}

\begin{figure}[t]
    \begin{center}
        \noindent\includegraphics[width=0.8\textwidth]{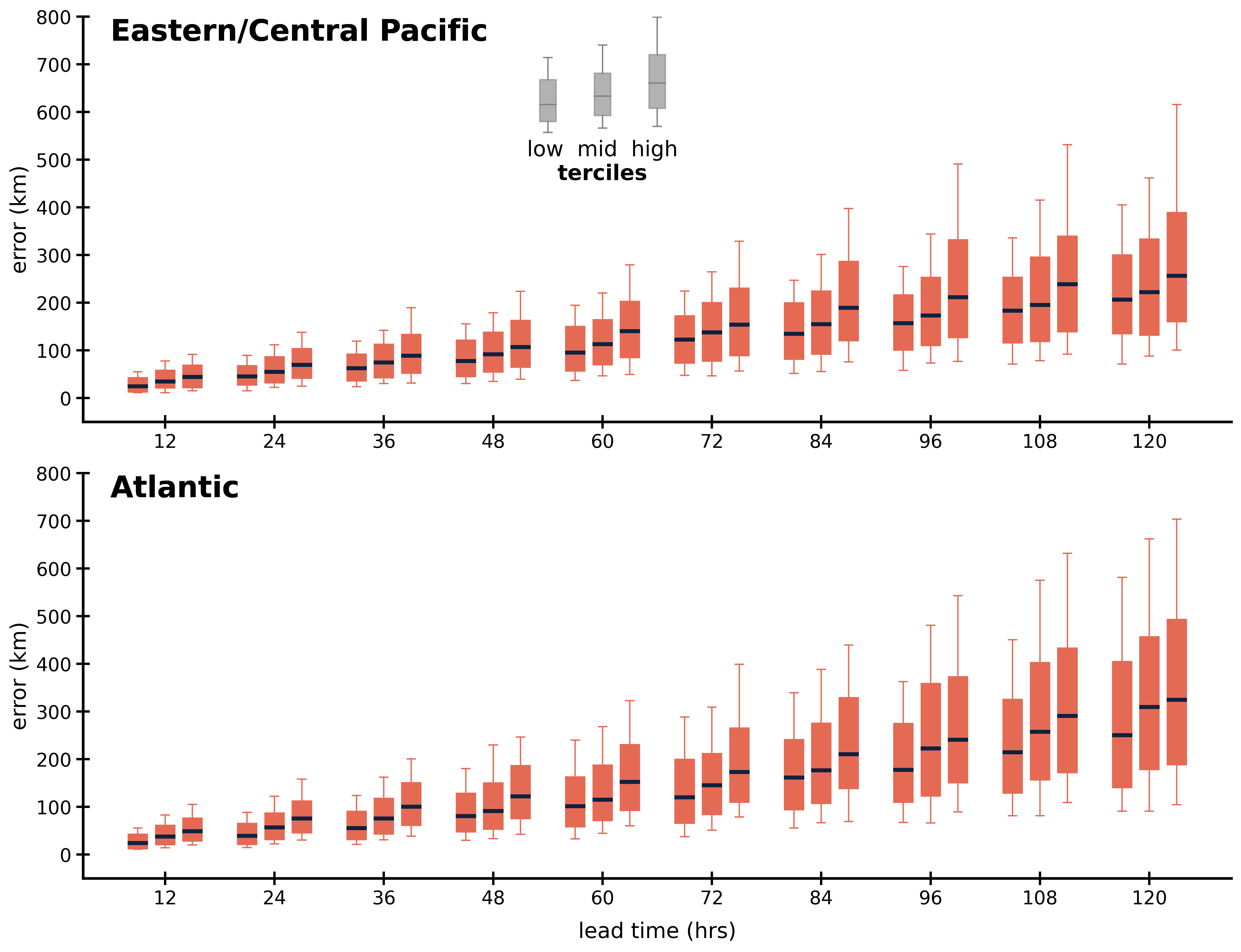}
        \caption{
            Interquartile range (IQR) versus error.
            Boxplots show the distribution of forecast error (filled area spans 25th to 75th percentile, whiskers out to 10th and 90th percentile) associated with the lower, middle, and upper tercile of IQR for each lead time.
            The IQR is a measure of the width of the predicted bivariate distribution.
                }\label{fig:iqr_error}
    \end{center}
\end{figure}

Many metrics support determining the calibration and validity of probabilistic models \citep{Gneiting2007}, and here we showcase two such metrics: the interquartile range (IQR) versus error, and the probability integral transform \citep[PIT,][]{Dawid1984}.
Figure \ref{fig:iqr_error} shows the IQR versus true error (the labels used in training) for both the Eastern Pacific and the Atlantic basins.

IQR values are computed as the difference between the 75th and 25th percentiles for each predicted bivariate normal, and are thus a measure of the width of each predicted distribution.
In Figure \ref{fig:iqr_error}, the IQR is divided in three bins for each lead time (lead time indicated along the horizontal axis): the lower, middle, and upper terciles of IQR for the set of predictions for that basin and lead time.
The true errors associated with each of these bins are shown, with the median (solid line), the 25th to 75th percentile (filled), and 10th and 90th percentile (whiskers) all indicated.
For well-calibrated networks, we expect the median error and the error range to be larger for larger IQR.
This is evident for each lead time in Figure \ref{fig:iqr_error}, where the distribution shifts to higher error as we move from the lowest IQR tercile up to the highest IQR tercile.
Static error distribution parameters such as those used in the cone of uncertainty are not able to capture this variability other than the basin and lead time dependence.

\begin{figure}[t]
    \begin{center}
        \noindent\includegraphics[width=0.8\textwidth]{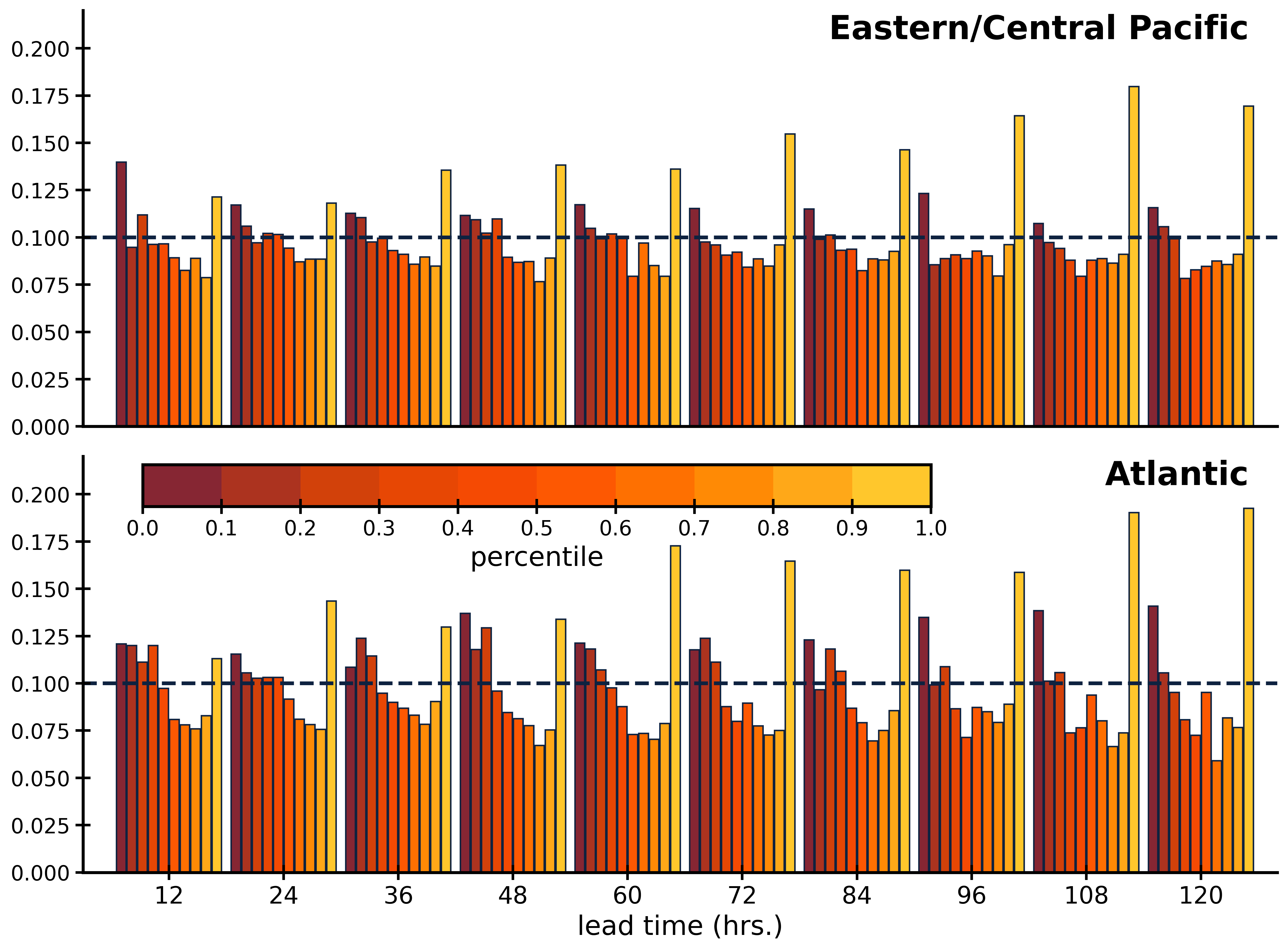}
        \caption{
            Probability Integral Transform (PIT) histogram for the Eastern Pacific (top panel) and Atlantic (bottom panel) for all forecast lead times.
            This metric describes how often the truth falls into each decile of the predictions (10th, 20th, 30th, etc.).
            A perfectly calibrated probabilistic model would have a uniform distribution of 0.1.
                }\label{fig:pit}
    \end{center}
\end{figure}

PIT histograms, shown in Figure \ref{fig:pit}, quantify how often the truth falls into a certain percentile of the predicted bivariate normal distribution's CDF.
PIT values are shown for every 10 percent increment, e.g., the leftmost bar for each lead time shows the fraction of the time the truth falls between the 0th and 10th percentile, the next bar is for the 10th to 20th percentile, and so on.
A perfectly calibrated model would be uniform with a constant value of 0.1, indicated in the figure by a dashed horizontal line.
For the Eastern/Central Pacific, our networks are making too many wide and narrow predictions (the rightmost and leftmost bars for each lead time are larger than 0.1).
The Atlantic shows the same, but slightly stronger, bias as the Eastern Pacific.
However, the values for most of the other bins are not too far from 0.1.

One way of quantifying how well calibrated our predictions are is to compare the PIT-D statistic, which measures the deviation of our PIT histogram from a uniform distribution, to the expected deviation.
The PIT-D statistic is given by $D = \sqrt{1/B\sum_k\left( b_k - 1/B\right)}$, while the expected deviation is given by $E[D] = \sqrt{(1-1/B)/(T \times B)}$, where $B$ is the number of bins, $T$ is the number of samples, and $k$ indicates summation over each bin \citep{NipenStull2011,Bourdin2014}.
Our predictions range from $D = 0.011$ up to $D = 0.037$, while the expected deviation is between $E[D] = 0.005$ and $E[D] = 0.010$.

\section{Results}\label{sec:results}

Satisfied that our framework produces reasonable and well-calibrated uncertainty predictions, we turn to the use of these predictions.
In particular, we analyze our predictions for all forecasts made by the NHC from 2013 through 2023 in the Eastern/Central Pacific and Atlantic basins.
We do this by using a leave-one-year-out method; we train our network on all but one year, then predict that left-out year.
We iterate through each left-out year to obtain predictions for every forecast without the network seeing that year in its training.

\begin{figure}[t]
    \begin{center}
        \noindent\includegraphics[width=0.9\textwidth]{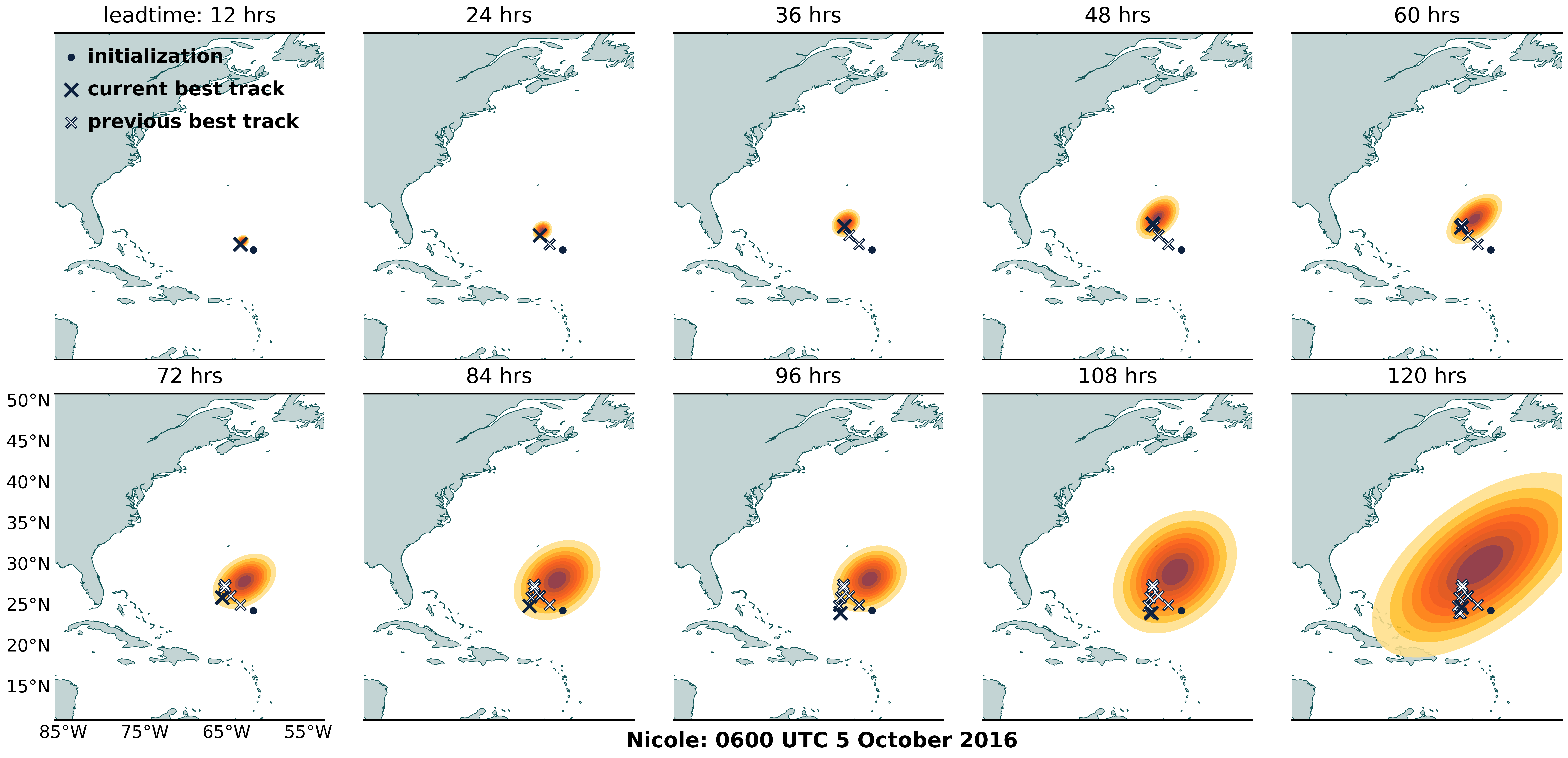}\\
        \noindent\includegraphics[width=0.9\textwidth]{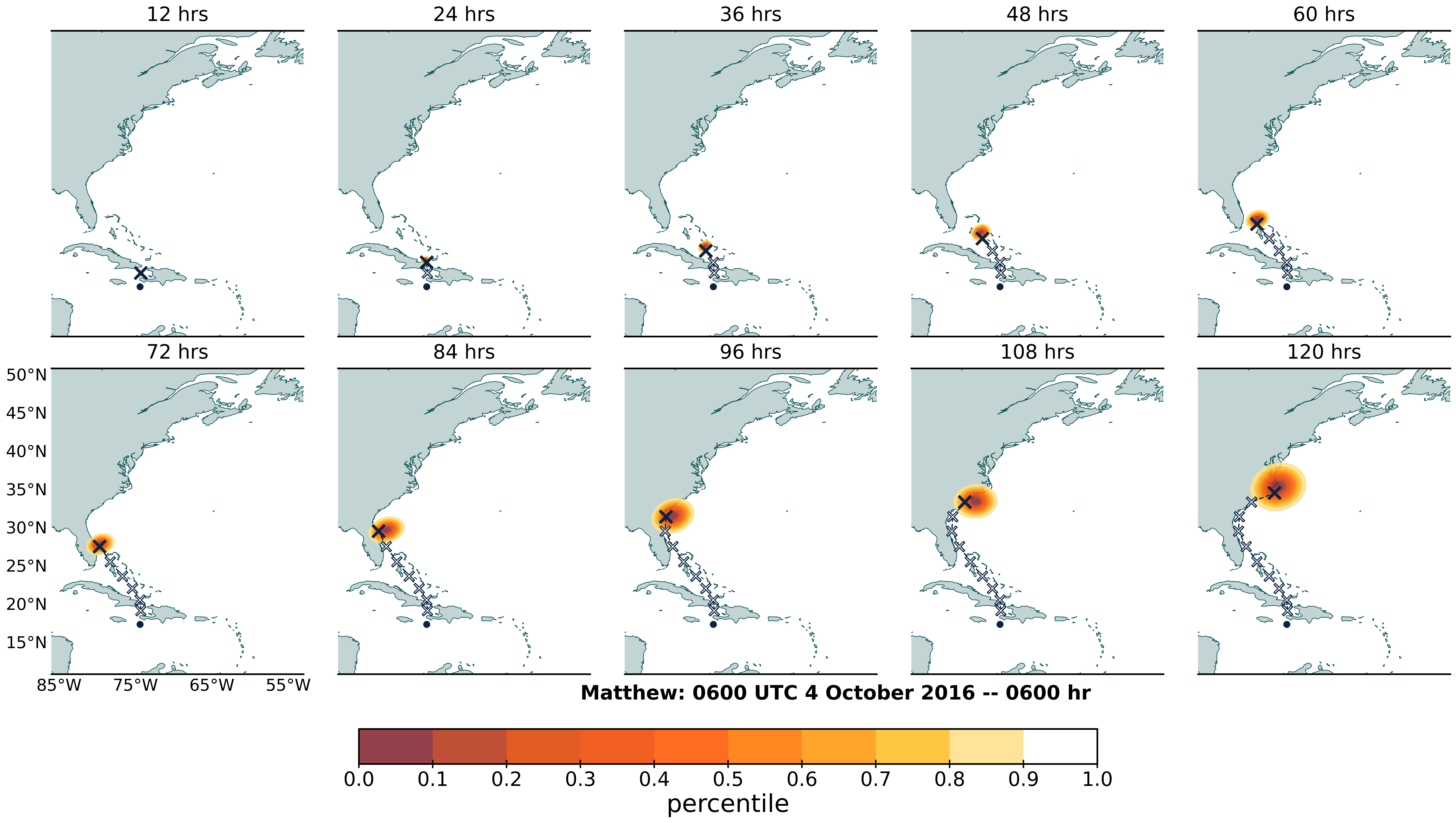}
        \caption{
            Two examples of our predictions, as described in Figure \ref{fig:prediction_schematic}.
            The initialization time is fixed and the forecast every 12 hours out to five days is shown, along with the best-track reconstruction.
                }\label{fig:storm_tracks}
    \end{center}
\end{figure}

Figure \ref{fig:storm_tracks} shows two examples of forecasts with our predicted bivariate normal CDF overlaid in the red-to-yellow shading.
The top panels show the forecasts made for hurricane Nicole at 0600 UTC 5 October 2016, and the bottom panels show the forecasts made for hurricane Matthew at 0600 UTC 4 October 2016.
In both cases, forecasts are shown out to five days.
The bivariate normal CDF is centered at the NHC official forecast location; we do not fit the location parameters ($\mu_x$ and $\mu_y$) of our bivariate normal, as described in Section \ref{sec:methods}.
We show two concurrent storms in the same basin to emphasize that our method predicts uncertainties based on forecast-specific inputs.
The larger predicted uncertainties for hurricane Nicole reflect that it was difficult to forecast.
Hurricane Matthew was easier to forecast with smaller error, also reflected in our predicted uncertainties.

Among the forecasts there are several examples that highlight the usefulness of the correlation parameter (i.e., the flexibility of our predicted bivariate normal shape).
This is especially apparent for Nicole, where the forecast was consistenly to the northeast of the truth.
Our predicted bivariates point in a northeast-southwest direction, emphasizing that the uncertainy is larger along that axis.

\subsection{Comparison with NHC Cone, GPCE Radii, and GEFS}\label{sec:nhc_cone}

Comparing our probabilistic estimate of track uncertainty directly to the NHC cone or GPCE radii is difficult; the cone and GPCE only provide a single radii value at each forecast time, but our method estimates the full error distribution.
However, one metric that can be used is the continuous ranked probability score (CRPS) \citep{Gneiting2007}.
The CRPS collapses to the mean absolute error when both parts are deterministic (both the prediction and truth CDFs are step functions), so it can be thought of as an extension to mean absolute error that allows for a probabilistic component.

For our two-dimensional case, we calculate the one-dimensional CRPS along both the latitude and longitude and multiply these, only integrating over the quadrant in which the truth falls.
For the NHC cone or GPCE, which are symmetric (circular), no further adjustments are necessary.
To account for the variable shape of bivariate normal predictions, the prediction CDF used to calculate the CRPS is the distribution conditioned on the truth along the other axis, e.g., to calculate the CRPS along the latitude axis, we condition on the true longitude error.
This is further explained in the Supplementary section S1.

\begin{figure}[!ht]
    \begin{center}
        \noindent\includegraphics[width=\textwidth]{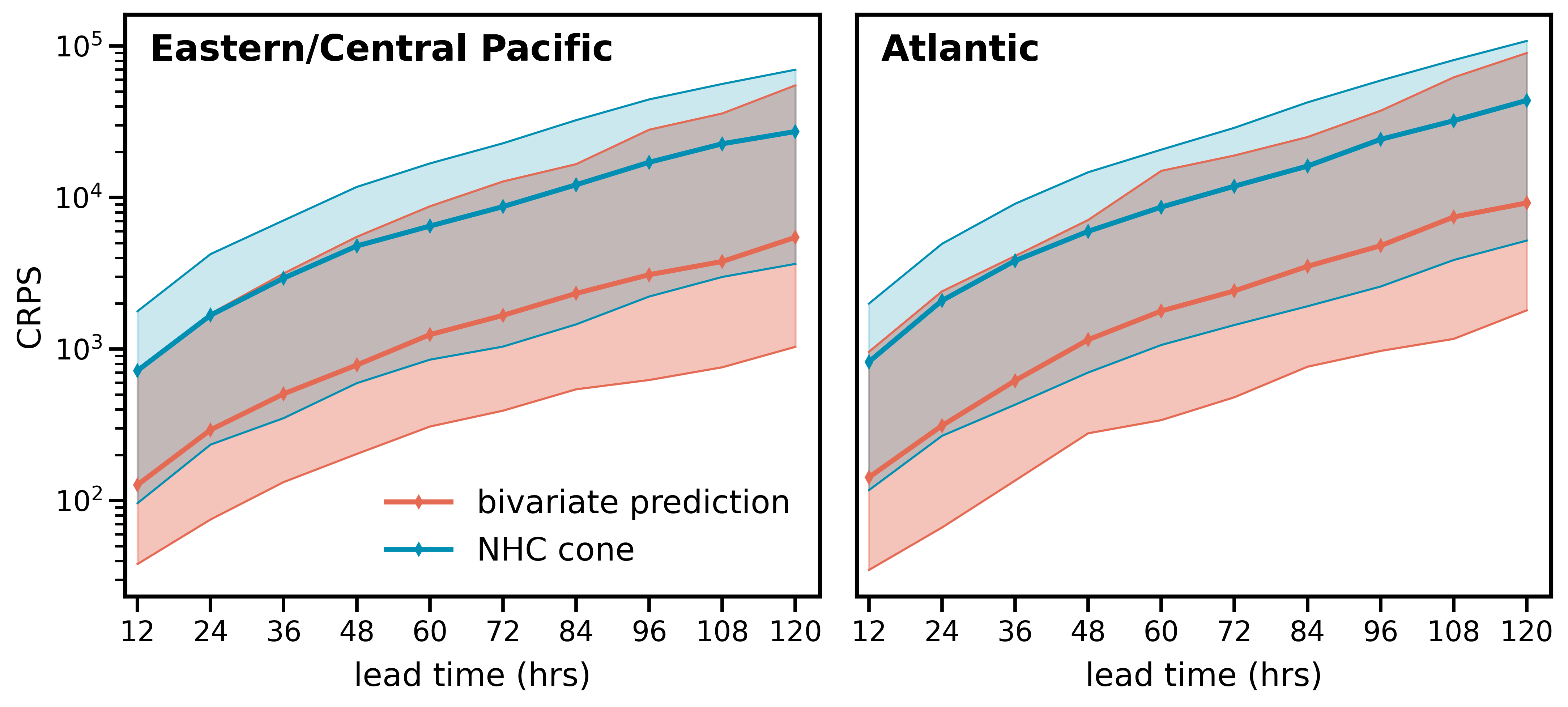}
        \caption{
            Continuous Ranked Probability Score (CRPS) for the NHC cone and the bivariate predictions as a function of lead time.
            The median for each is shown as the solid lines, while the shaded area encloses the $10$th to $90$th percentile of CRPS values.
            A lower value of CRPS indicates a better prediction (a CRPS value of zero indicates a perfect prediction, with the entirety of the prediction weight at the truth, e.g., a delta function).
                }\label{fig:crps}
    \end{center}
\end{figure}

Figure \ref{fig:crps} shows the results of the CRPS calculation for our predictions (red) and the NHC cone (blue).
As with mean absolute error, a lower CRPS value is better.
The CRPS is calculated for all forecasts, and the median for these is shown as a solid line, with the 10th to 90th percentiles shaded.
According to the CRPS metric, our predictions are a better estimate of the true error than the NHC cone for a majority of the forecasts.
This is unsurprising, as the NHC cone is static throughout a season and has a fixed circular symmetry.

We calculated the CRPS for the GPCE predictions as well and found these to be very similar to the NHC cone.
Using the standard deviation of the GEFS member displacements from the GEFS mean, and the corresponding correlation, we construct bivariate normal predictions and calculate CRPS for GEFS.
We find the GEFS CRPS to be very similar to our bivariate predictions.
We reiterate that running an ensemble such as GEFS has a much higher computational cost than the method presented here.
Both GPCE and GEFS CRPS are presented in Supplementary section S2.

We can also make a comparison by choosing a specific percentile of our distribution and comparing only the associated ellipse, though this undermines one of the main strengths of our predictions, to the NHC cone or GPCE.
With this inhibited version of our prediction, we can look at the binary question of whether each of the predictions captures the truth.

\begin{figure}[!ht]
    \begin{center}
        \noindent\includegraphics[width=\textwidth]{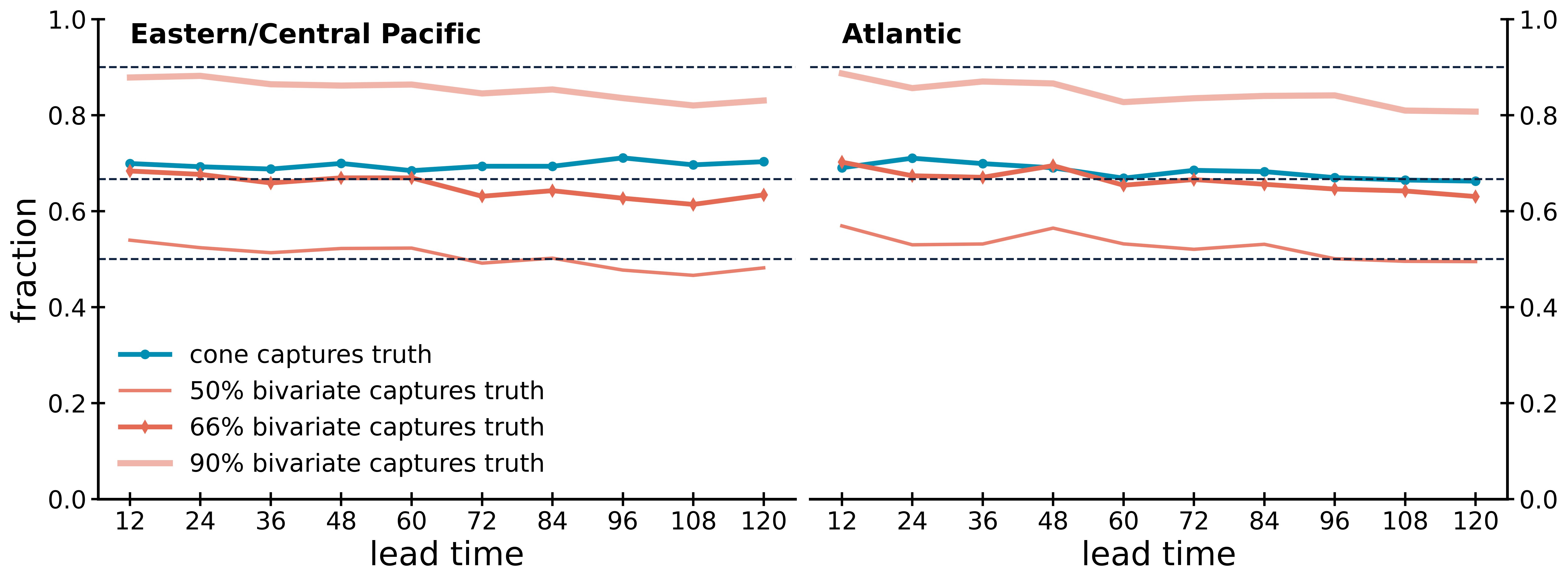}
        \caption{
            Fraction of cases where the bivariate predictions (red) and the NHC cone (blue) capture the truth as a function of lead time.
            Shown are several percentiles for the bivariate; from thinnest to thickest, the $50$th, $66$th, and $90$th percentile ellipses, respectively.
            This emphasizes the strength of using the distribution for prediction and allows for both probabilistic and tailored deterministic predictions (e.g., minimizing misses or minimizing false alarms).
                }\label{fig:capture_fraction}
    \end{center}
\end{figure}

Figure \ref{fig:capture_fraction} shows the fraction of forecasts where the NHC cone captured the true TC location (blue), and the fraction of forecasts where several of our percentile ellipses (red) captured the truth.
We can use any percentile from our predictions, but we show only three: the 50th, 66th, and 90th percentile ellipse capture fractions.
The dashed lines show perfect calibration, e.g., the 50th percentile of our distributions capture the truth 50 percent of the time.
The 66th percentile ellipse in Figure \ref{fig:capture_fraction} remains remarkably close to the perfect 66 percent dashed line.
This supports the results in Figure \ref{fig:pit}, but additionally emphasizes the flexibility of our method.
Specifically, our method allows for a subjective choice of either minimizing misses (using a higher percentile ellipse, or setting a higher percentile threshold), or minimizing false alarms (using a lower percentile ellipse, or a lower percentile threshold).

\subsection{Landfall Events}\label{sec:landfall}

\begin{figure}[!ht]
    \begin{center}
        \noindent\includegraphics[width=\textwidth]{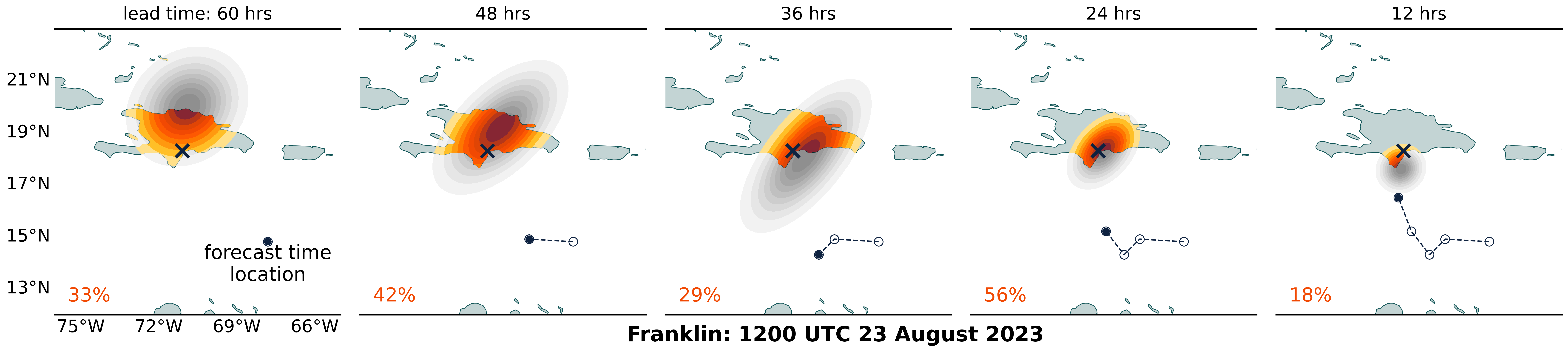}\\ 
        \noindent\includegraphics[width=\textwidth]{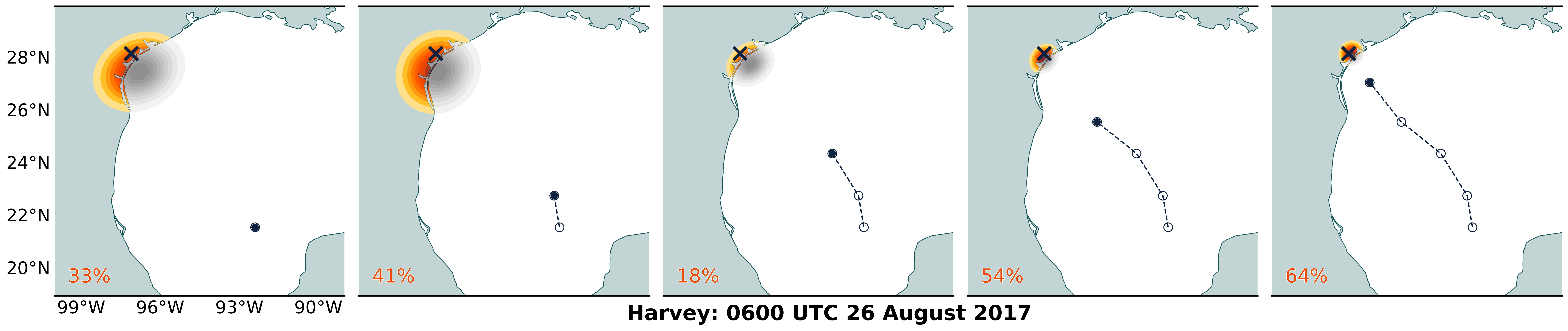}\\ 
        \noindent\includegraphics[width=\textwidth]{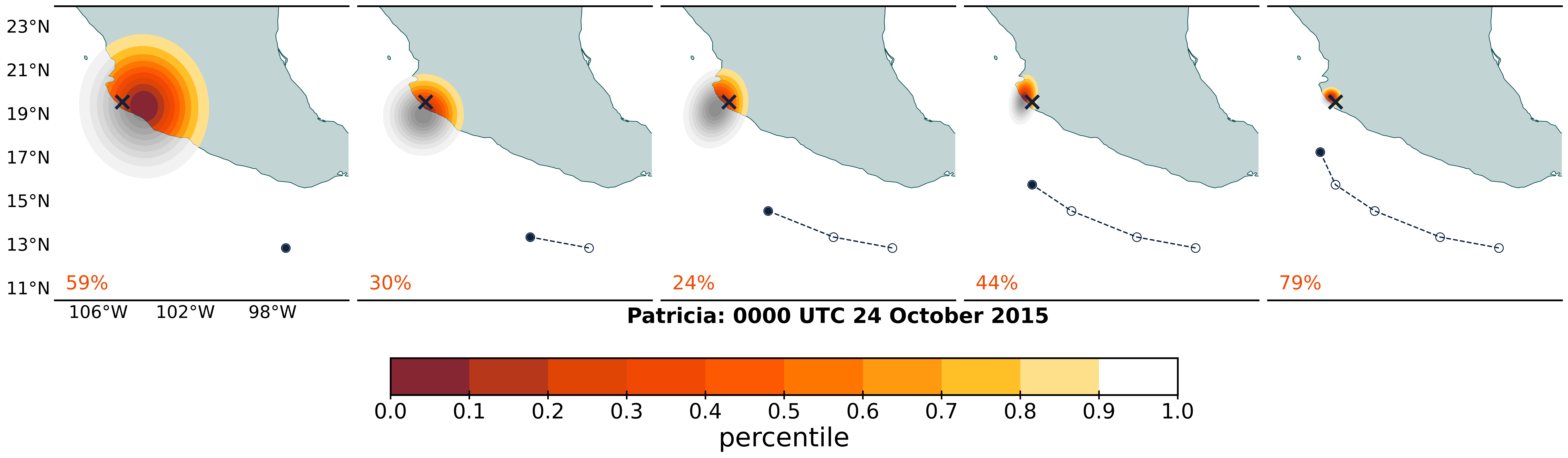} 
        \caption{
            Three examples of predictions for TCs as they make landfall.
            For each storm, the nearest synoptic time to landfall, and the corresponding location, is used.
            The forecasted time is held fixed (panels go from longer to shorter lead time).
            Landfall location is indicated by a cross and forecast initialization location is indicated by a solid dot (unfilled dots for previous times).
            Predictions are shaded red-to-yellow over land and grey-scale over ocean.
            The distribution integrated over land gives a probability for the TC to make landfall.
            The probability (as a percent chance) is shown in the lower left corner of each panel.
                }\label{fig:landfall_tracks}
    \end{center}
\end{figure}

Using our method, we can make probabilistic statements about landfall by integrating the portion of our predicted uncertainty that is over land at each forecast to obtain a probability of the TC making landfall at that time.
Several examples of this are shown in Figure \ref{fig:landfall_tracks}.
The top panel shows hurricane Franklin (2023) making landfall over the Dominican Republic; the middle panel shows hurricane Harvey (2017) making landfall over Texas in the United States; and the bottom panel shows hurricane Patricia (2015) making landfall over Jalisco in Mexico.
As expected, the uncertainty decreases as we approach the forecasted time.

We divide all forecasts into cases where the TC did make landfall and cases where it did not.
Across all forecasts, there were 1,889 instances where a TC made landfall in the Atlantic, and 478 instances where a TC made landfall in the Eastern/Central Pacific.
We checked the calibration metrics (PIT, IQR versus error) for this subsample of predictions and found that the predictions remained well calibrated (not shown).

\begin{figure}[!ht]
    \begin{center}
        \noindent\includegraphics[width=1.0\textwidth]{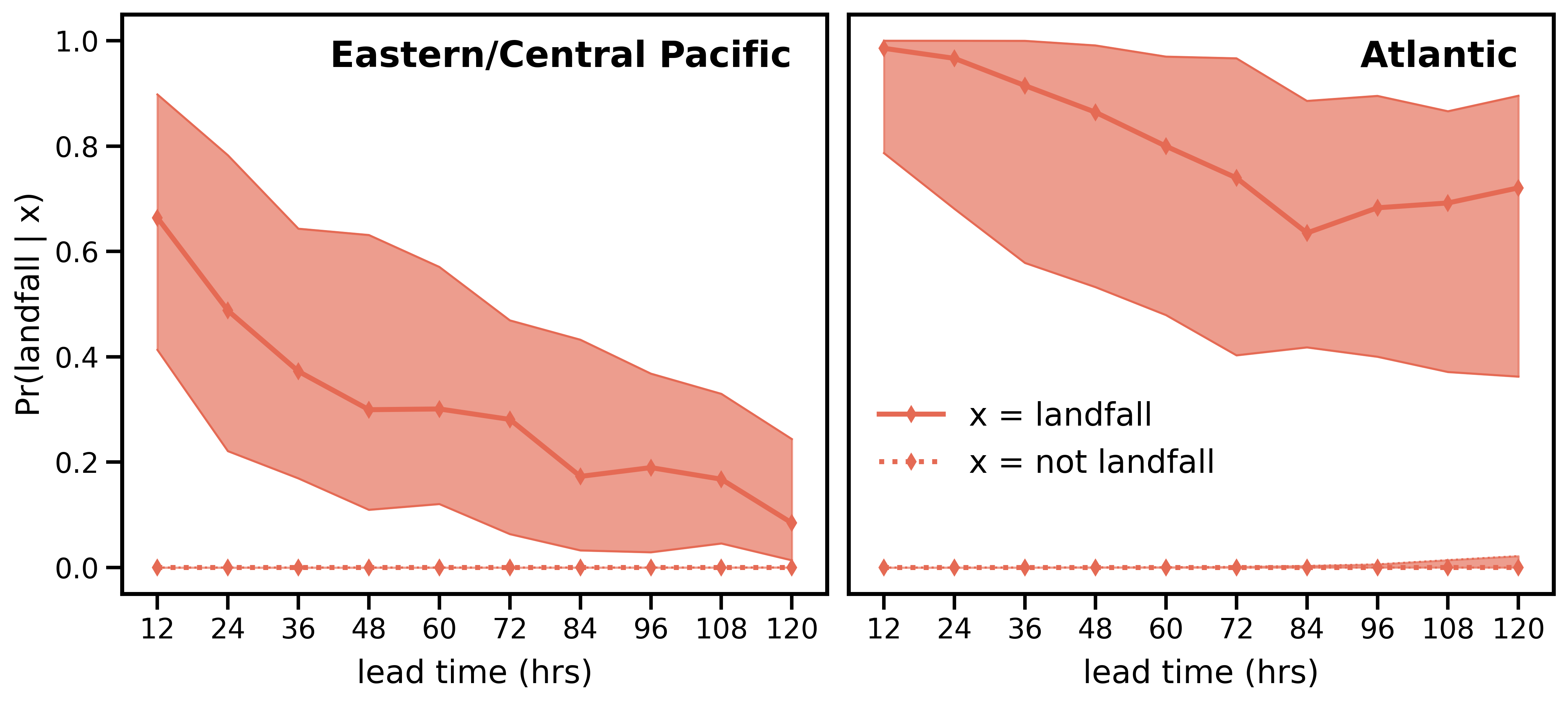}
        \caption{
            Probability of landfall.
            The panels show the probability of landfall (bivariate integrated over land, see Figure \ref{fig:landfall_tracks}) for TCs in the Eastern/Central Pacific (left) and Atlantic (right) as a function of lead time.
            The solid lines show the median probability for cases when the TC did make landfall, prob(landfall $|$ landfall), and the dotted lines show the median probability for cases when the TC did not make landfall, prob(landfall $|$ not landfall).
            The shaded areas indicate the $25$th to $75$th percentile of the distribution of landfall probabilities.
                }\label{fig:landfall_stats}
    \end{center}
\end{figure}

Figure \ref{fig:landfall_stats} shows the probability of landfall for both cases; when there was landfall, $P(\text{landfall }|\text{ landfall})$, and when landfall did not occur $P(\text{landfall }|\text{ no landfall})$.
While the probability of landfall for cases when it did not occur is very low for both basins, the probability of landfall for cases when there was landfall look fairly different.
In the Atlantic, our predicted landfall probability is high for all lead times, with the median remaining above the $0.5$ line throughout.
The Eastern/Central Pacific probabilities decay strongly with lead time, likely due to the very small sample size available for training, a result of fewer landmasses in the path of Eastern/Central Pacific TCs.
For example, at 120 hours there are only 19 forecasts where a TC made landfall in the Eastern/Central Pacific.

We repeat the preceding analyses for early forecasts (capture fraction, CRPS, landfall probability) and find similar performance in all cases.
These are shown in the Supplementary section S3.
We apply the preceding analysis (CRPS) to Atlantic landfall cases to compare our bivariate predictions to the NHC cone.
We find a large improvement over the NHC cone, with the mean NHC cone CRPS approximately twice as large (worse) than our bivariate predictions.
This is shown in Supplementary section S6.

\section{Conclusion}\label{sec:conclusion}

We have developed and tested a method of estimating tropical cyclone track uncertainty.
Using forecast-specific inputs and the true forecast error as the label, we train a neural network to predict the parameters of a bivariate normal distribution.
The distribution serves as our estimate of the TC track uncertainty for that forecast.
The network is trained on a dataset from the NHC and CPHC, which includes 11 years (2013 through 2023) of forecasts, 10 lead times (12 to 120 hours), and the Atlantic and combined Eastern and Central Pacific basins.
The loss used in training the network is the negative log probability of the truth (difference between the forecast location and the best-track reconstruction), given the predicted distribution.

We have shown that predictions using our method are well calibrated using the probability integral transform (PIT) metric.
We have also compared the interquartile range (IQR) of our predictions to the true forecast errors.
According to the continuous ranked probability score (CRPS), our method produces better uncertainty estimates than the NHC cone and the GPCE track uncertainty estimates, and is comparable to predictions from the Global Ensemble Forecast System (GEFS).
The probabilistic nature of our predictions allows for a subjective, expert-based decision on whether to emphasize minimizing false alarms or minimizing misses.
We have also shown that a probabilistic approach can be used to robustly estimate the probability of landfall events.

The move toward a probabilistic estimate of track uncertainty is already a priority for forecasting centers \citep{Dunion2023,Conroy2023}.
Currently, the NHC (and many other operational forecast centers) estimates TC track forecast errors using historical errors of their operational forecasts from the previous five years.
These are static (same for the entire season, circularly symmetric) and deterministic (a single-valued uncertainty estimate).
Our method produces uncertainty estimates that are dynamic (forecast-specific, variable shapes) and probabilistic.

Once trained, the computational cost of predictions using our method is negligible, potentially making it more appealing than costly ensemble methods that may or may not have enough members for the distribution to converge.
The method is flexible so that new models can be included provided an adequate training sample is available.
For example, the HWRF model used in this study is being replaced in NHC forecasts by the HAFS model \citep{Hazelton2021}, which will require a different set of track models to be used as input.
In addition, parameters from ensemble forecast systems such as ensemble spread can be added as input to the neural network.
Thus, our method is a strong candidate to improve operational track uncertainty estimates.

\clearpage
\acknowledgments
This work was supported by NOAA grant NA22OAR4590525.


\datastatement
Operational model outputs, environmental variables, and best track data were obtained from the National Hurricane Center.
The code and data set used in this work can be found at \url{https://github.com/mafern/tcane_track}, and will be given a permanent DOI on Zenodo at the time of publication.


\bibliographystyle{ametsocV6}
\bibliography{references}

\clearpage
\pagenumbering{gobble}


\section*{Supplementary material}

# S1   CRPS in Two Dimensions

The CRPS metric is generally meant to compare a probabilistic distribution to a deterministic truth, but can also be used when both parts are deterministic (as is the case for the NHC cone and GPCE). The CRPS is essentially the area between the CDF of the prediction and the CDF of the truth, where the truth CDF is represented by a step function (zero inside the true value, one outside). In one-dimensional cases, the CRPS is given by

$$\text{CRPS} = \int \left(F_{\text{pred}}(x) - F_{\text{true}}(x)\right)^2 \, dx = \sum_i \left(F_{\text{pred}}(x_i) - F_{\text{true}}(x_i)\right)^2 \, \Delta x. \qquad (1)$$

The CRPS is well defined for one-dimensional distributions, but is more ambiguous for higher dimensional spaces. Specifically, there are several potential ways to represent the truth (deterministic component) and regions over which to integrate. For example, one can construct a step function in two dimensions, then integrate over the whole space or just the quadrant containing that step. [1] proposed the energy score, a multivariate extension to CRPS using expectation values of samples drawn from the prediction distribution. Alternatively, [2] defines the multivariate CRPS similarly to the univariate case, splitting the integral and integrating over the multivariate space. However, [3] note that the multivariate CRPS does not fully account for correlations. Specifically, [3] show that misspecified correlations between the true distribution and the predicted distribution are not punished appropriately.

In the likely case that TC track forecast errors in latitude and longitude are not equal, but have some correlation, the above methods of calculating CRPS would artificially punish our bivariate predictions. Instead, we calculate the CRPS for each dimension separately and take the product of these as the two-dimensional CRPS. This still accounts for the correlation between the two dimensions.

Figure S1 illustrates how we calculate the CRPS. For each potential true value (square and cross), the resulting one-dimensional CDFs for the prediction and the truth are shown for latitude and longitude. The bivariate normal is a better prediction for the truth represented by the cross, and this is reflected in the CRPS values, which are 0.16 for the cross, and 0.70 for the square.

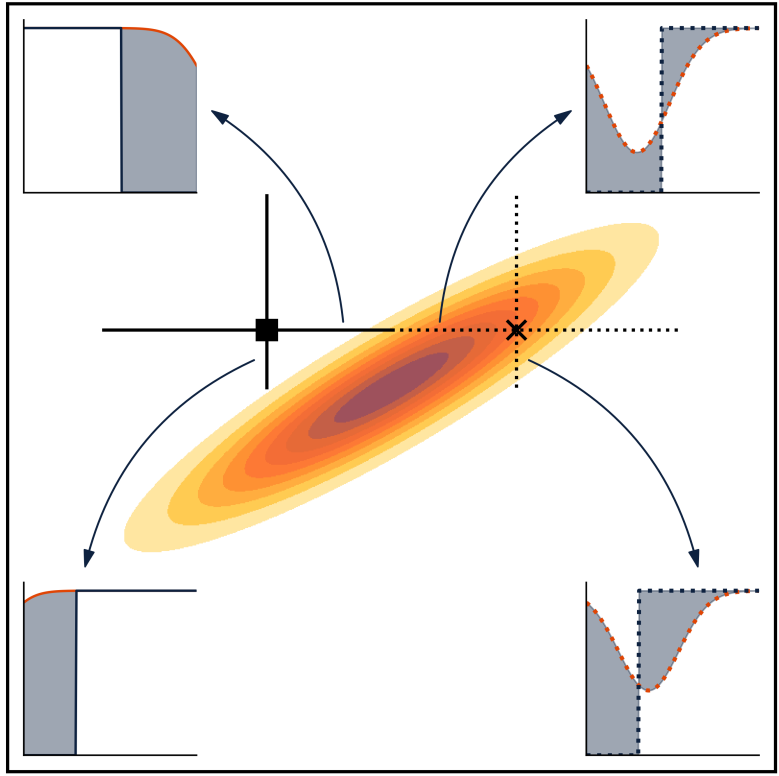

Figure S1: Calculating two-dimensional CRPS from the two one-dimensional CRPS values. An example bivariate normal prediction is shown in red-to-yellow shading, and two potential truths that are mirrors (across the vertical) of each other are shown to illustate the reason for using the conditional distributions.

## S2 Comparison to GPCE and GEFS

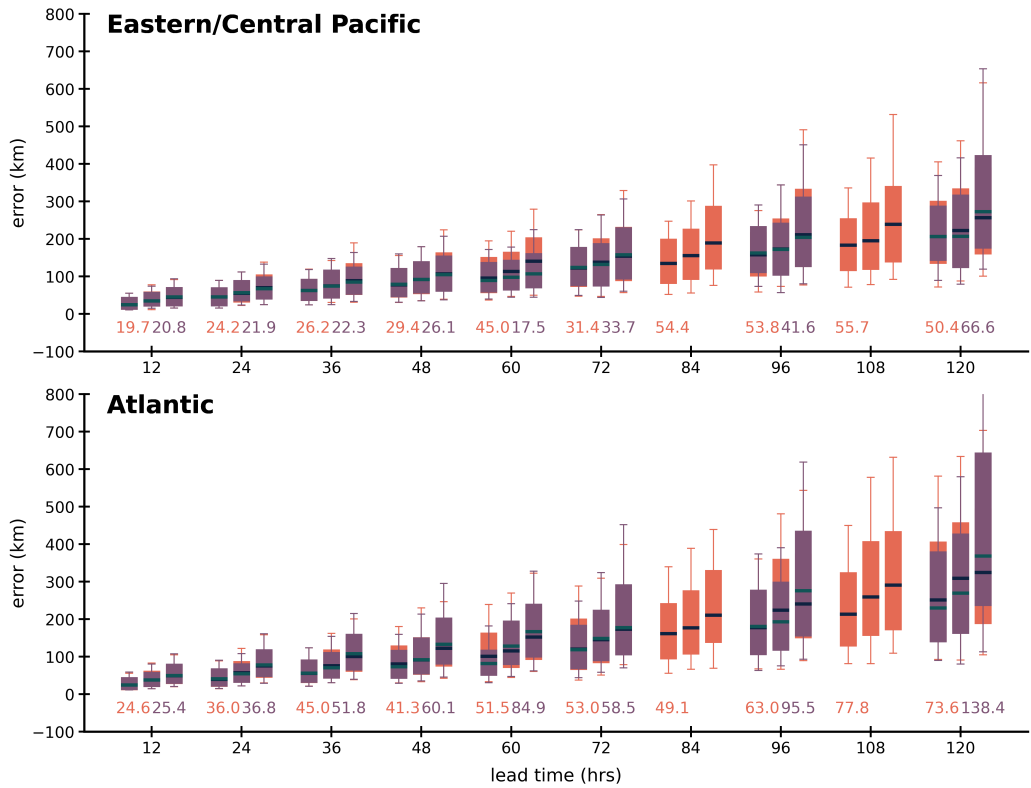

Figure S2: Interquartile range (IQR) versus error. Boxplots representing the GPCE method are plotted on top of boxplots representing our method. Values for each lead time are the difference between the highest and lowest tercile medians for our method (left) and GPCE (right). GPCE was not available for lead times of 84 and 108 hours.

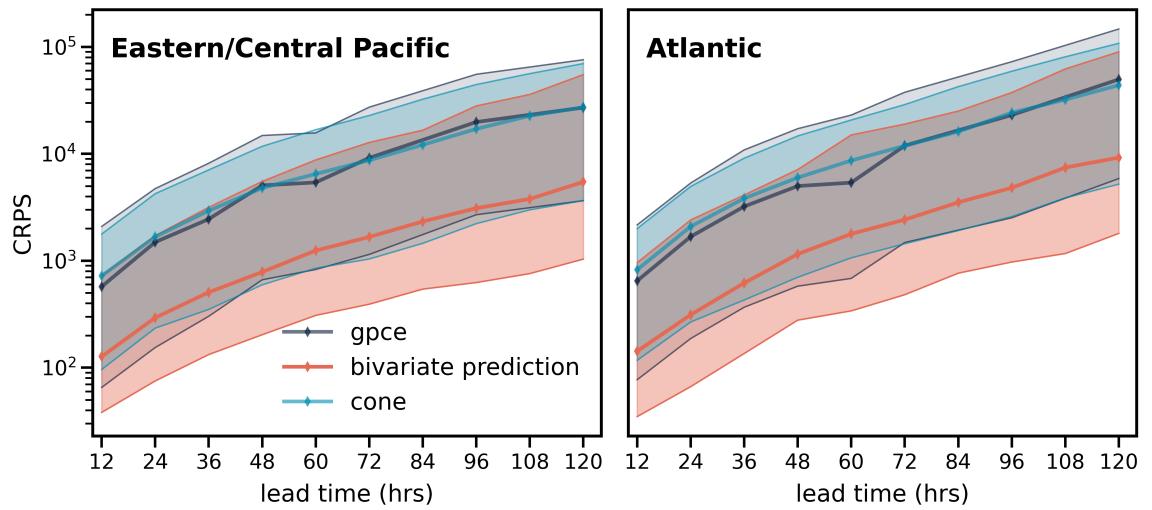

Figure S3: Continuous Ranked Probability Score (CRPS) for the NHC cone, our bivariate predictions, and GPCE, as a function of lead time. GPCE was not available for lead times of 84 and 108 hours. The NHC cone and GPCE are very similar in both median (solid lines) and 10th to 90th percentile range (shaded area).

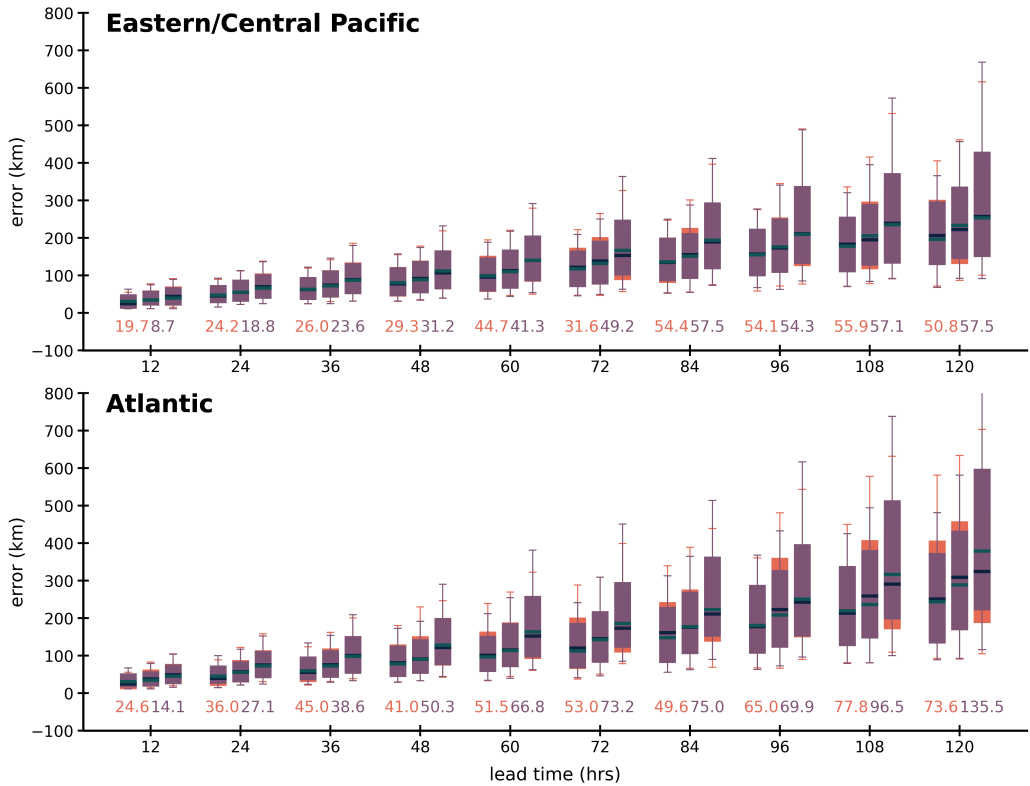

Figure S4: Same as Figure S2, but for GEFS. The GEFS contained a maximum of 20 ensemble members from 2013 through 2019, then the number was increased to a maximum of 30 members in 2020. In our sample, we have 9955 forecasts with 30 members and 28101 forecasts with 20 members. The remaining forecasts have between 3 and 29 members.

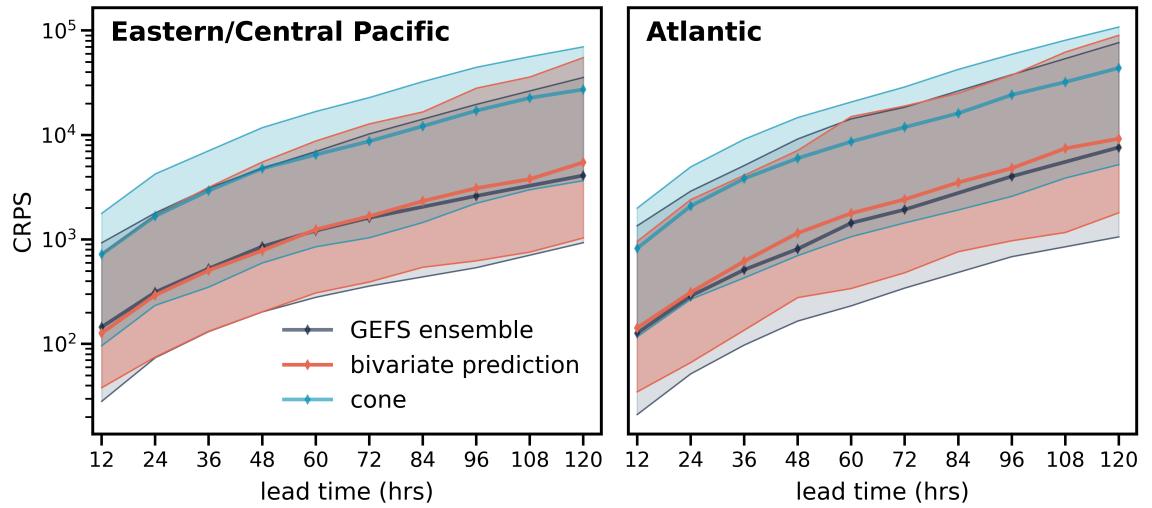

Figure S5: Same as Figure S3, but for GEFS. See note on ensemble size of GEFS in Figure S4 caption. Our bivariate predictions and GEFS are very similar in both median (solid lines) and 10th to 90th percentile range (shaded area).

## S3  Early Forecasts

During the NHC forecasting cycle, early forecasts are made based on the consensus (model average) and these are subsequently used to inform the official NHC forecast. Our method of predicting track uncertainty may be useful for this early forecast, so we repeat all of the analyses previously shown for the official forecast.

The IQR versus error (Figure S6) shows the same trend with tercile and lead time as the late forecasts. The PIT histogram (Figure S7) is also very similar to the late forecast. The PIT-D statistic for these early forecasts ranges from $D = 0.009$ up to $D = 0.032$, while the expected deviation is between $E[D] = 0.005$ and $E[D] = 0.010$.

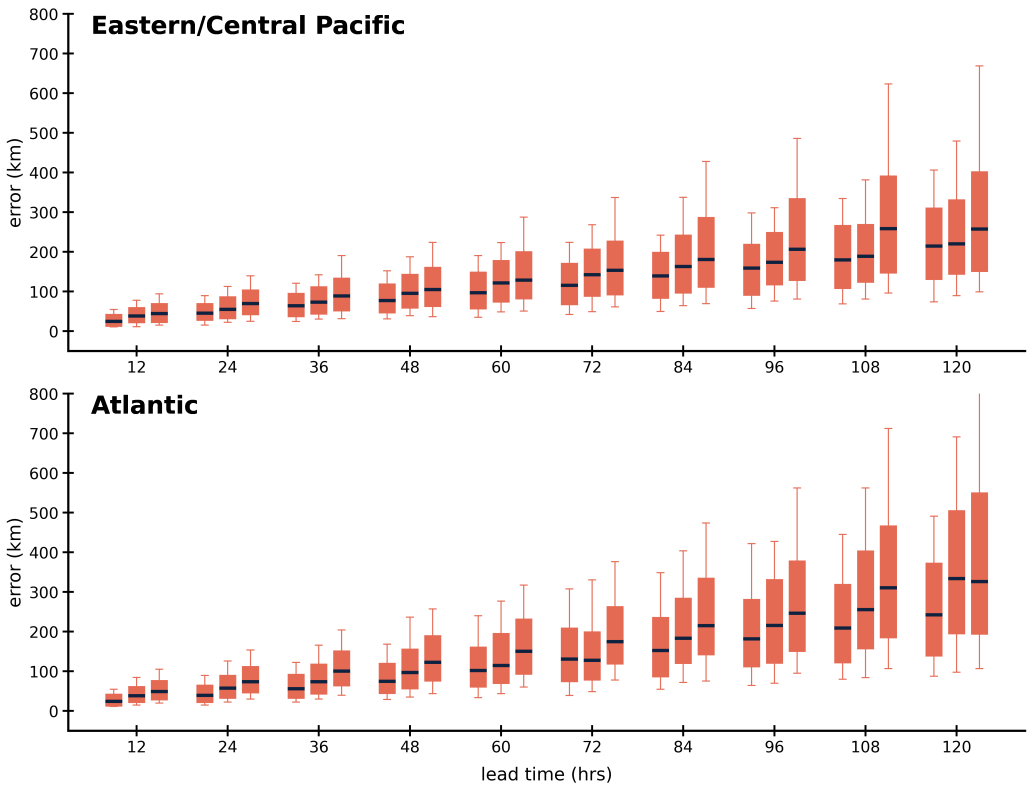

Figure S6: Interquartile range (IQR) versus error for early forecasts.

The capture fraction (Figure S8) for early forecasts is very similar to the late forecasts. The probability of landfall (Figure S9) for cases where landfall did occur is slightly lower for the early forecasts. On average, landfall probabilities for landfall cases are 0.08 lower for the Eastern Pacific, and 0.18 lower for the Atlantic.

Figure S10 shows the CRPS for the early forecasts. The CRPS values are marginally better (lower) for the early forecasts in both the Eastern Pacific and Atlantic, but are qualitatively

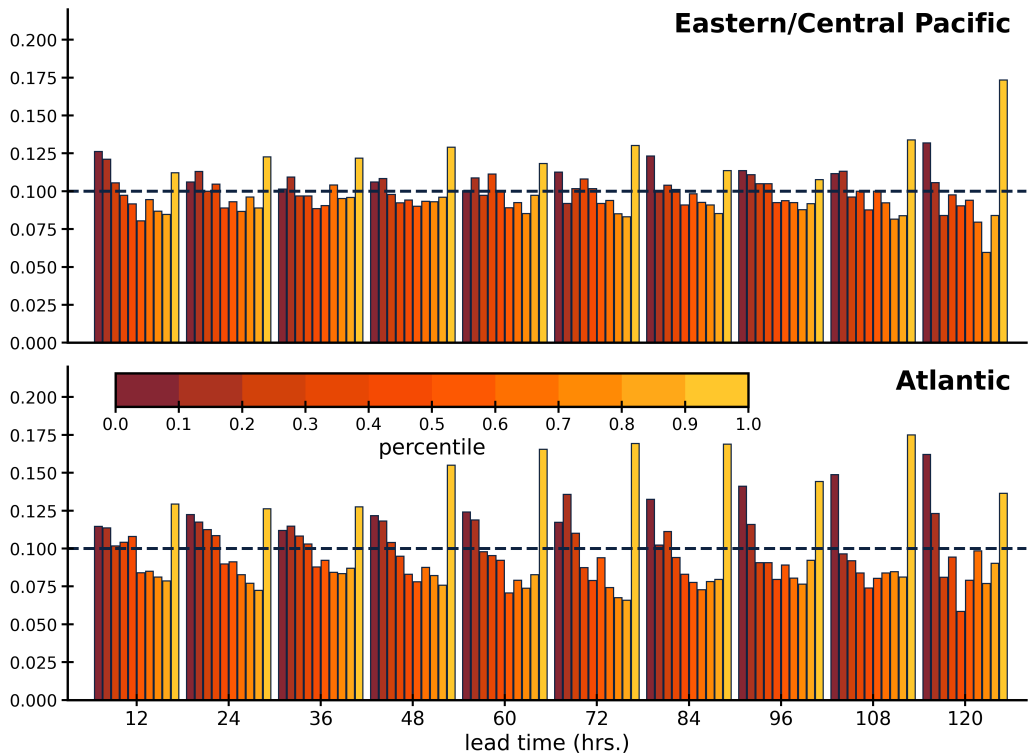

Figure S7: PIT histogram for early forecasts.

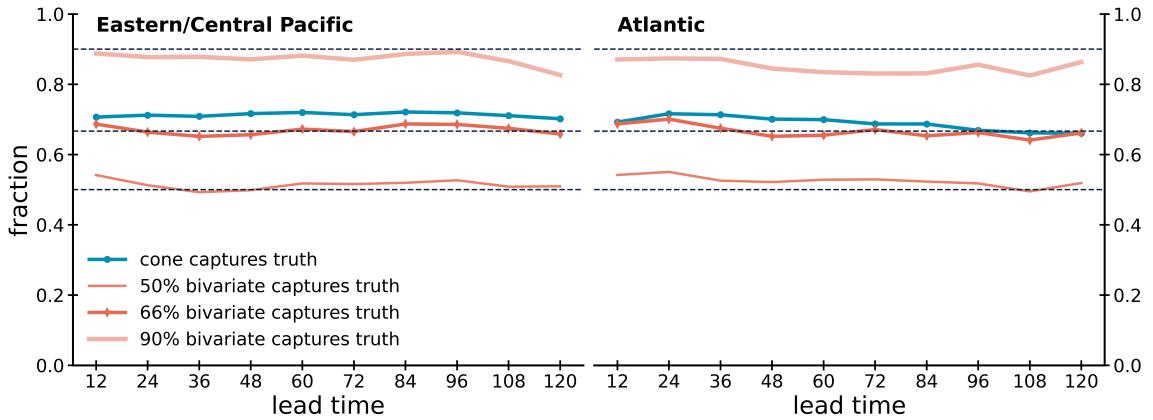

Figure S8: Capture fraction for early forecasts

unchanged.

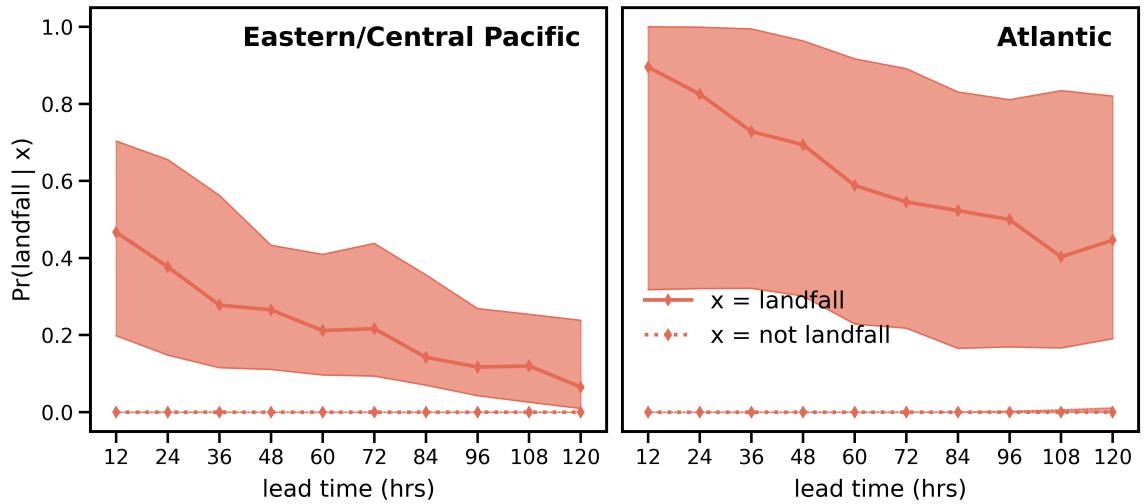

Figure S9: Landfall probability for early forecasts

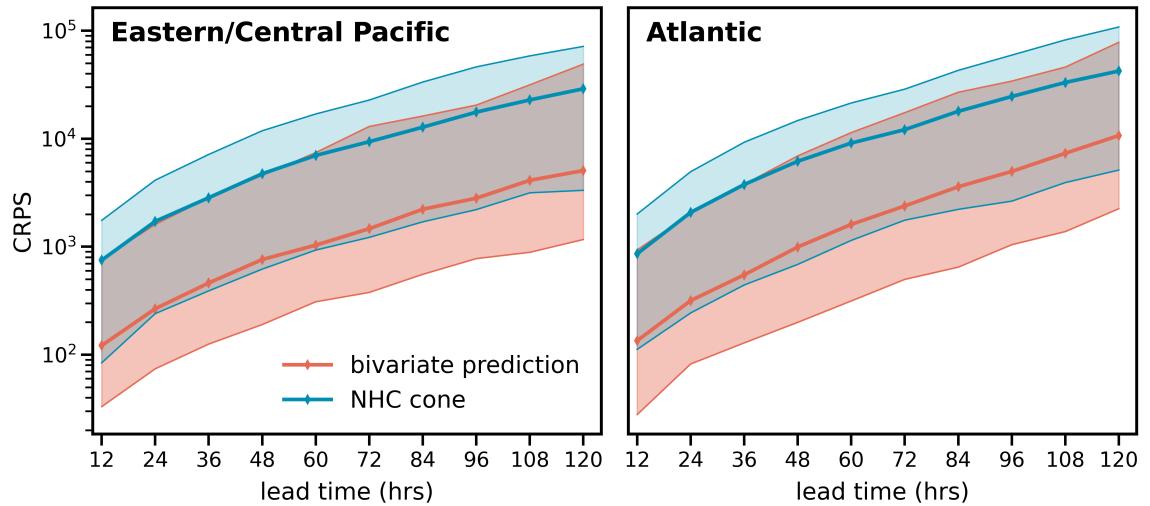

Figure S10: CRPS as a function of lead time for the NHC cone and the early forecast bivariate predictions. The median for each is shown as the solid lines, while the shaded area encloses the 10th to 90th percentile of CRPS values.

## S4 Explainable AI using SHAP

There are many explainable artificial intelligence (XAI) methods [4] aimed at understanding how opaque models came to a set of predictions. Here we use SHapley Additive exPlanations [SHAP; 5] to explore how the inputs affect the predictions, i.e., the feature relevance. We discuss explanations for a subset of inputs to the prediction of $\sigma_x$ for the Atlantic basin here; SHAP values for the remaining outputs for both basins are at the end of this section.

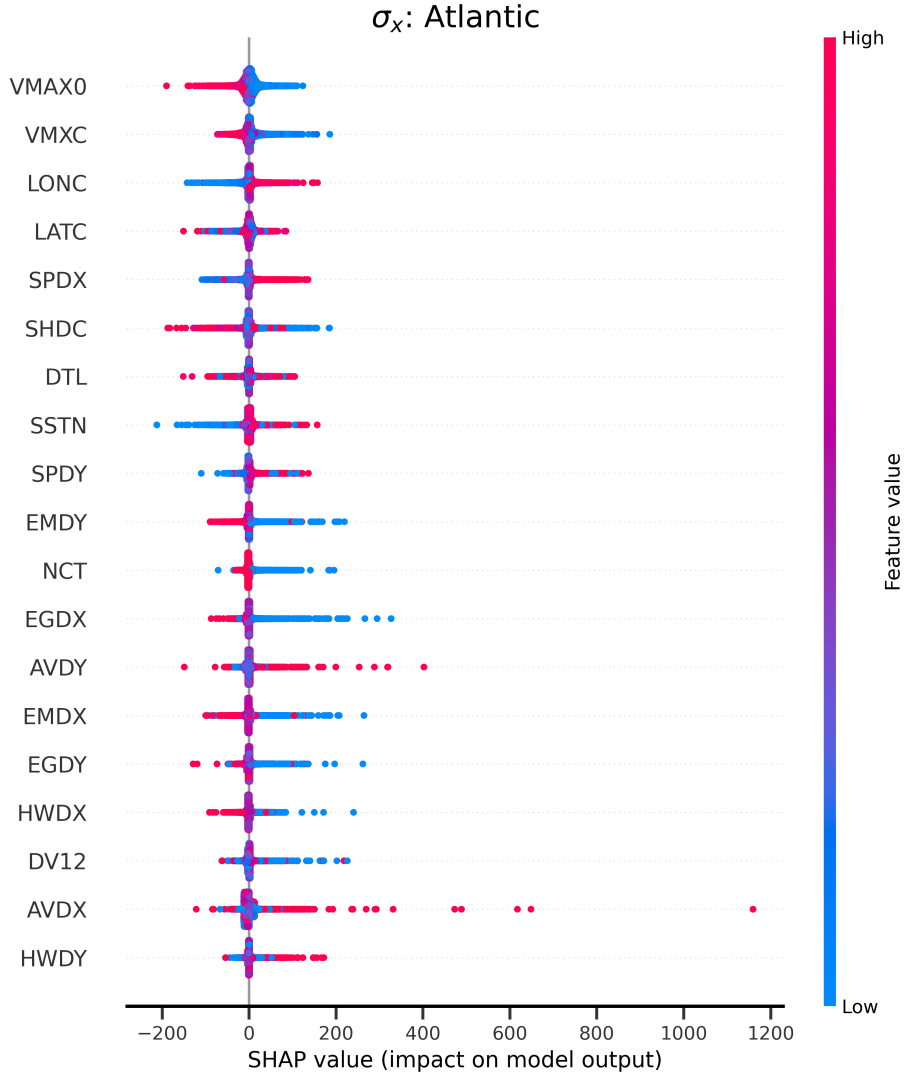

Figure S11: SHAP values for $\sigma_x$ in the Atlantic.

In Figure S11, input features are shown for $\sigma_x$ predictions in the Atlantic. The intensity feature (VMAX0) indicates that a higher intensity decreases the predicted uncertainty (smaller bivariate

widths). This may be due to high values indicating the storm is already well organized, potentially decreasing uncertainty on the storm's future intensity and thus its track. The two speeds (SPDY, SPDX) agree with each other; a larger speed in both the latitudinal and longitudinal directions increases the uncertainty. Our network predicts larger uncertainty for faster moving storms.

The consensus estimate for the latitude (LATC) is less cleanly separated, but generally a higher latitude increases the predicted uncertainty. The mean latitude value is near the edge of the tropics, so redder values are in the subtropics or mid-latitudes, while bluer values are in the tropics. There is some lead time dependence for LATC; at shorter lead times (less than 72 hours) a higher latitude actually decreases the predicted uncertainty, while at longer lead times a higher latitude increases the predicted uncertainty. This is consistent with the fact that some of the most difficult track forecasts are associated with the timing of re-curvature into the mid-latitude westerlies. If a storm is already at higher latitudes at short lead times, it is more likely that it already recurved and so is easier to predict than a storm at higher latitudes at longer lead times that may not have recurved at the start of the forecast. The consensus estimate for the longitude (LONC) indicates that more eastern longitudes increase the predicted uncertainty. Our network predicts larger uncertainty for storms that are farther from land.

The effect of sea surface temperature (SSTN) is for higher temperatures to increase the predicted uncertainty. This is likely due to the potential for a storm in a high SSTN environment to intensify. It may also indicate the storm is still forming. The effect of vertical shear (SHDC) on predictions is nuanced. In Figure S11, larger SHDC values (redder colors) both increase and decrease the predicted uncertainty, while low SHDC values (bluer colors) preferentially increase the predicted uncertainty. High shear may or may not disrupt the storm, but low shear allows the storms to intensify [6].

We discuss the effect on our predictions due to the UKMet global model (EGDY, EGDX) inputs. We chose the UKMet model because in the training set it was a middle performer, with track errors generally smaller than HWRF but larger than ECMWF and GFS. Inputs from these models are deviations from the consensus prediction, so high values indicate the model prediction is east or north of the consensus, while low values indicate the model prediction is west or south of the consensus. Since the consensus is the mean of the available models, deviations will necessarily cover both east and west, north and south (except in the case where all available models agree). For the UKMet model, a deviation east or north of the consensus leads to lower predicted uncertainty. This is consistent across the other three models for the longitude component, while for the latitude component the models are split (some indicate that deviations to the north lower predicted uncertainty, some raise the predicted uncertainty).

For most inputs the effect on the predictions is the same for the Eastern/Central Pacific basin. One notable difference is that the effect of LONC on our predictions is for higher LONC to lead to lower predicted uncertainty, a reversal from the Atlantic result, but consistent with LONC primarily indicating proximity to land.

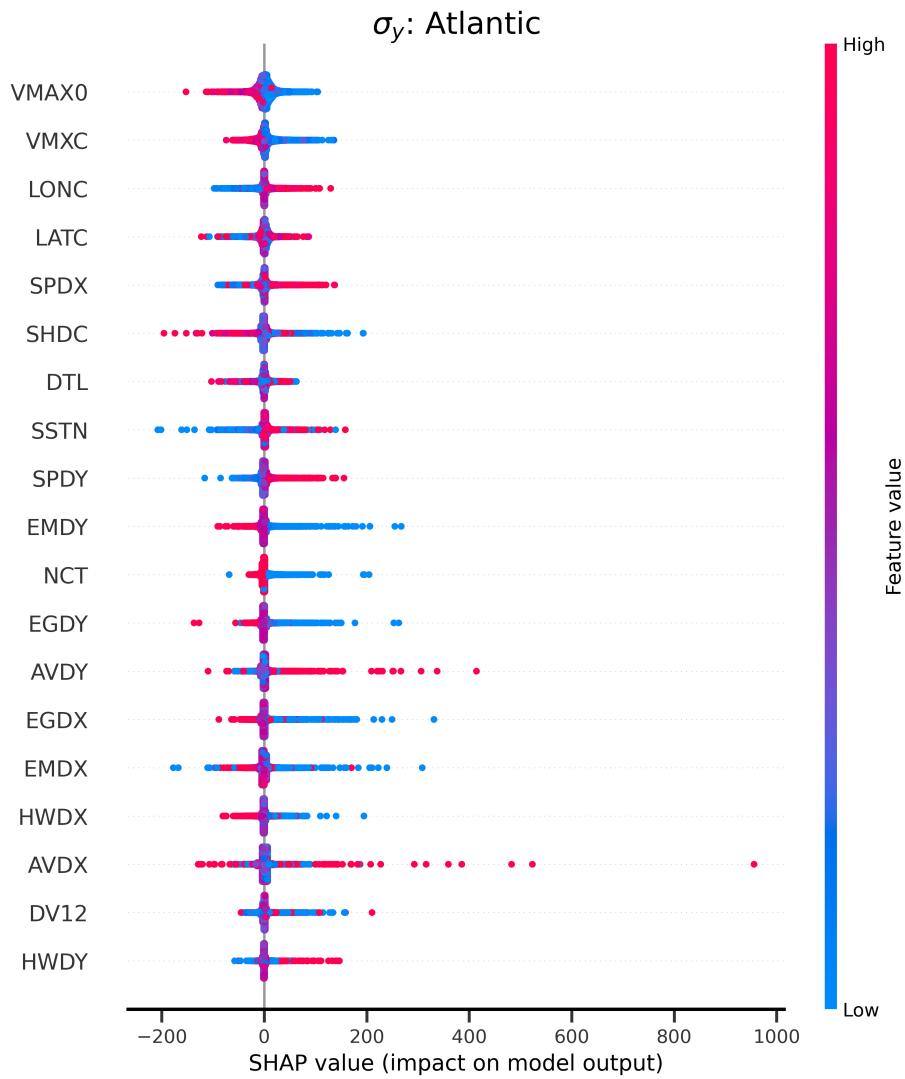

Figure S12: SHAP values for $\sigma_y$ in the Atlantic.

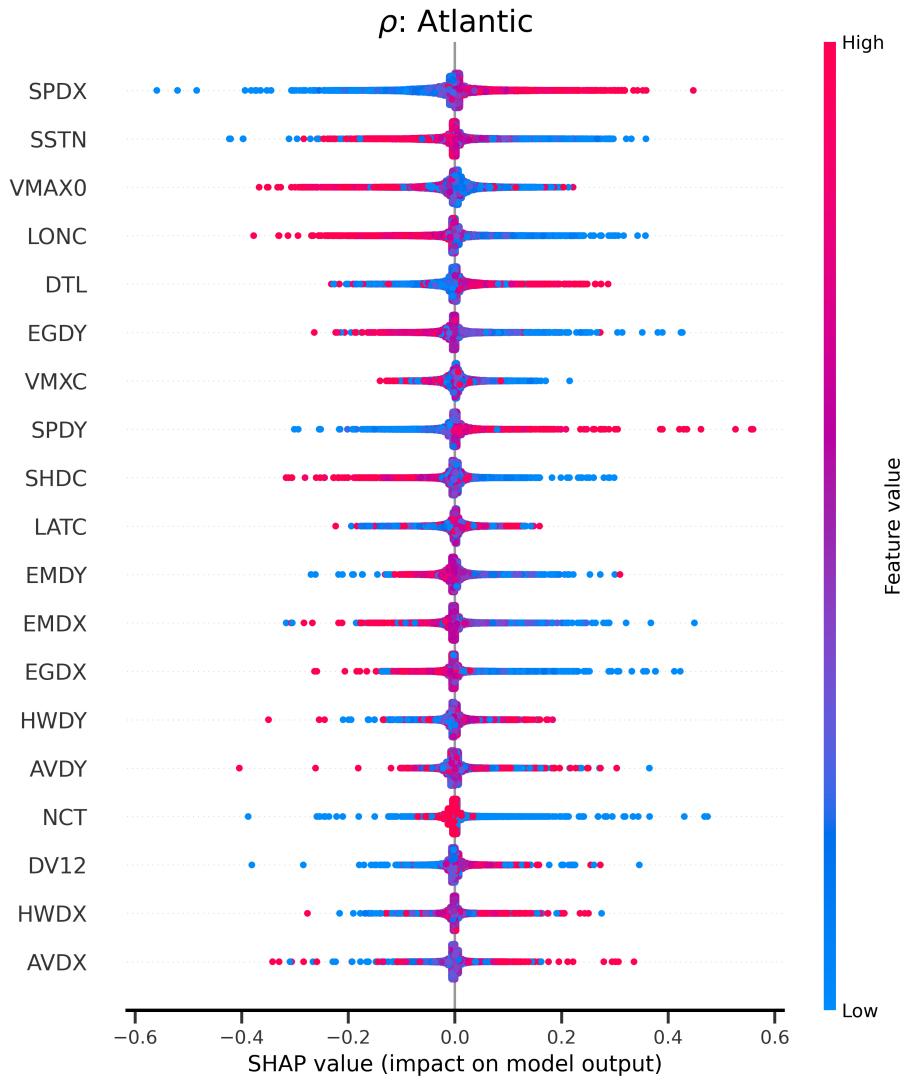

Figure S13: SHAP values for $\rho$ in the Atlantic.

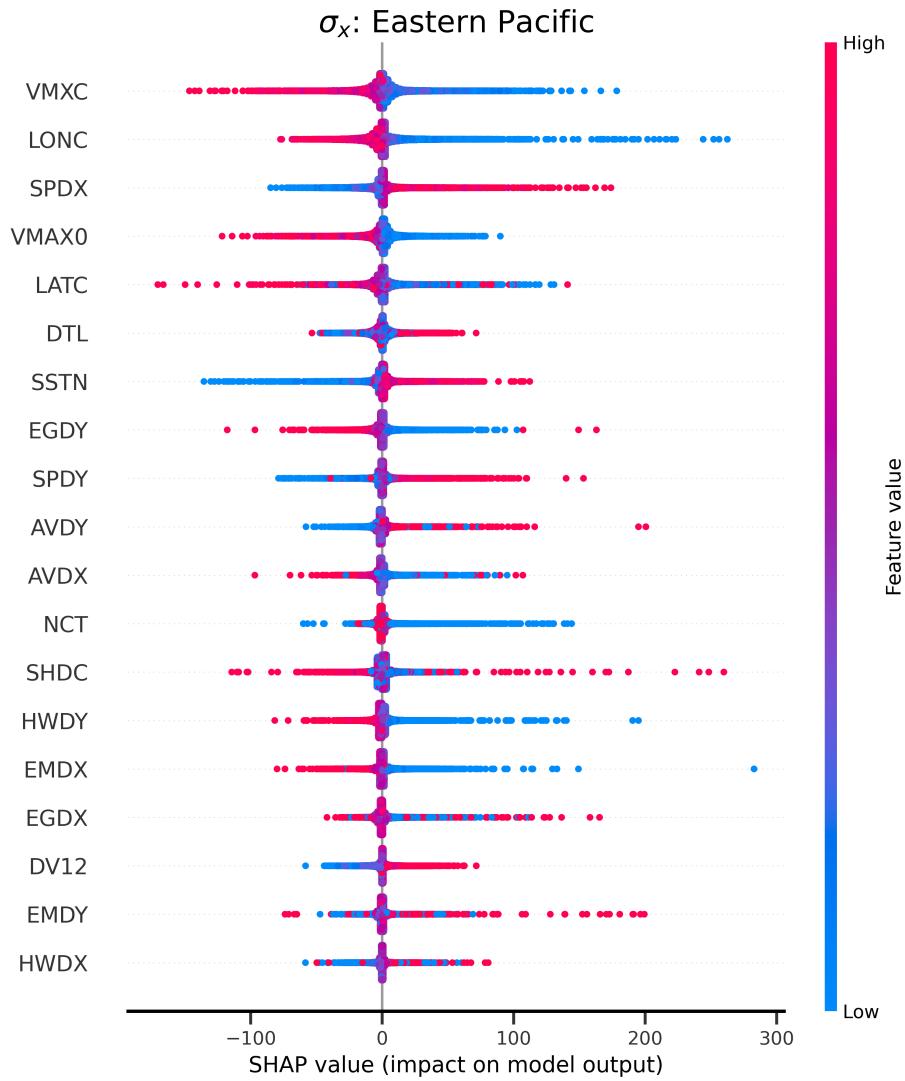

Figure S14: SHAP values for $\sigma_x$ in the Eastern/Central Pacific.

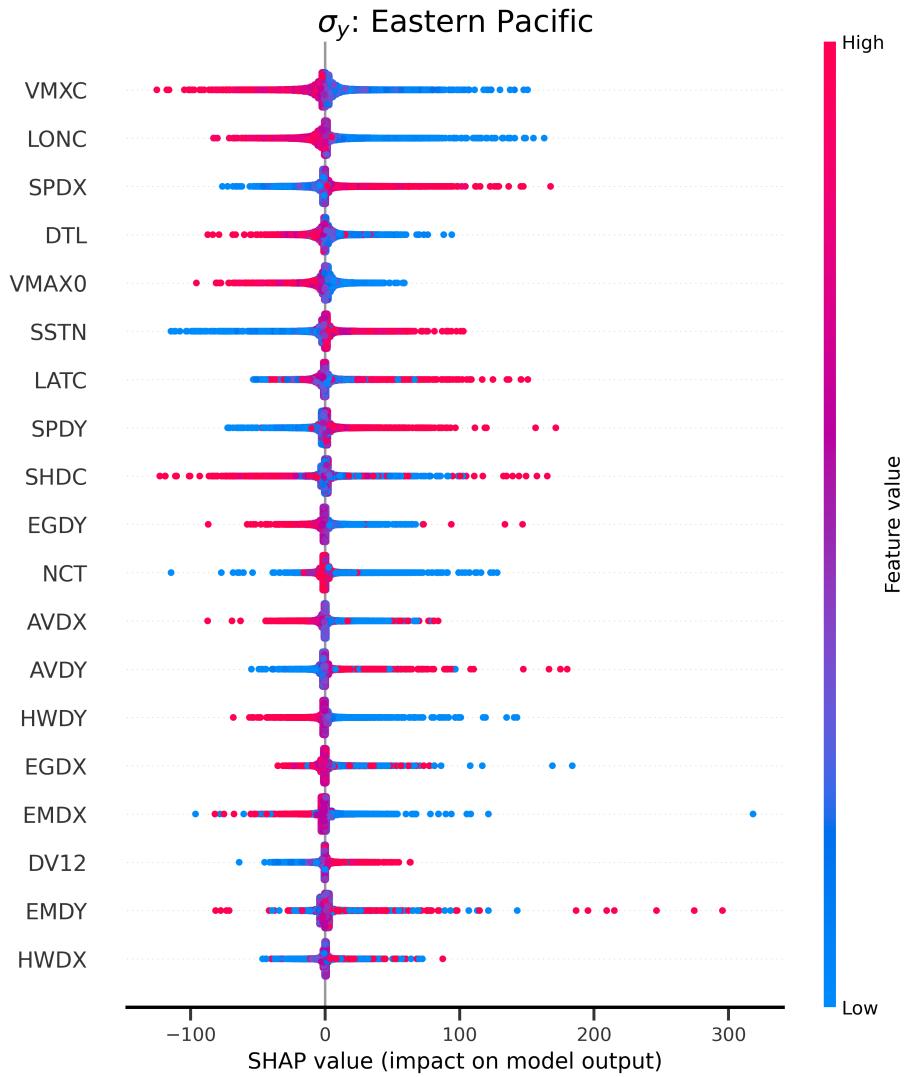

Figure S15: SHAP values for $\sigma_y$ in the Eastern/Central Pacific.

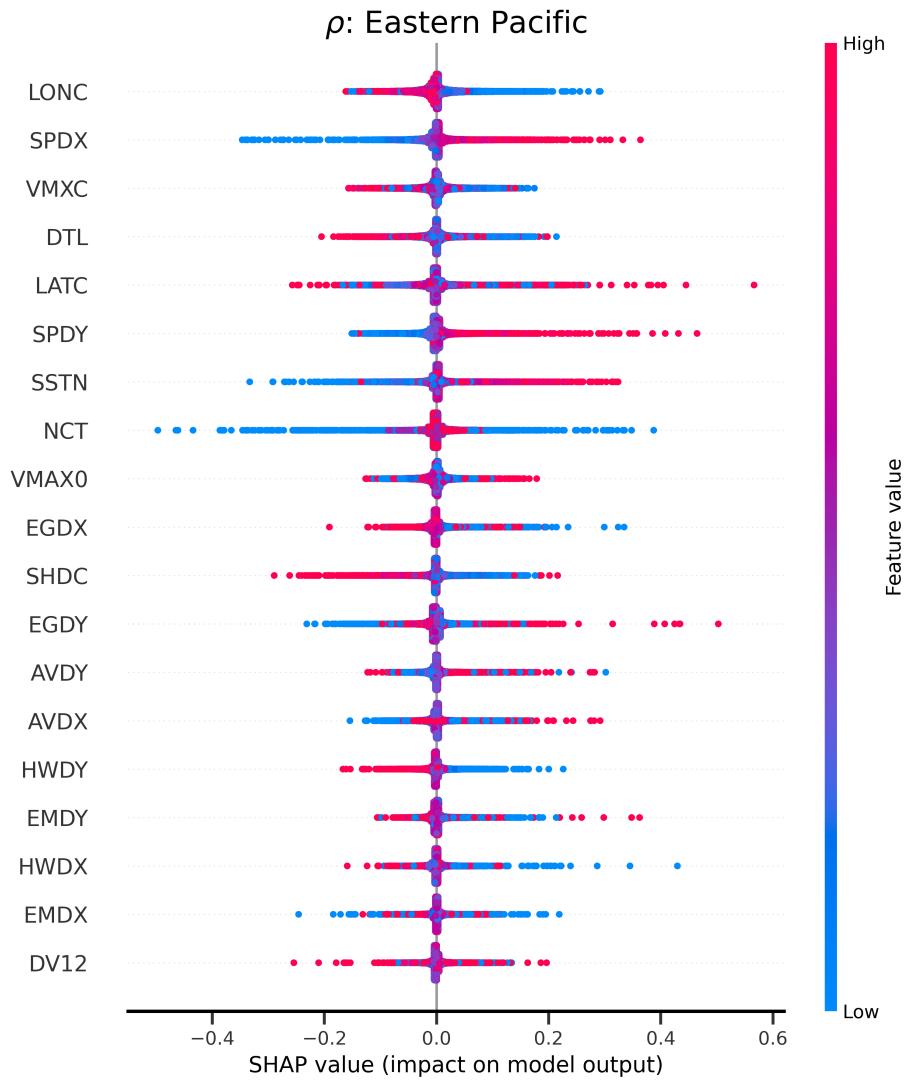

Figure S16: SHAP values for $\rho$ in the Eastern/Central Pacific.

# S5 NHC Cone of Uncertainty Values

Table 1: NHC Cone of Uncertainty values in nautical miles for the Atlantic basin for each year and lead time used in this study. For leadtimes 84 and 108, the earlier and later lead times are averaged. Rows are year, columns are lead times (in hours).

|      | 12 | 24 | 36 | 48 | 60  | 72  | 84  | 96  | 108 | 120 |
|------|----|----|----|----|-----|-----|-----|-----|-----|-----|
| 2013 | 33 | 52 | 72 | 92 | 110 | 128 | 152 | 177 | 203 | 229 |
| 2014 | 33 | 52 | 72 | 92 | 108 | 125 | 147 | 170 | 198 | 226 |
| 2015 | 32 | 52 | 71 | 90 | 106 | 122 | 146 | 170 | 197 | 225 |
| 2016 | 30 | 49 | 66 | 84 | 99  | 115 | 140 | 165 | 201 | 237 |
| 2017 | 29 | 45 | 63 | 78 | 92  | 107 | 133 | 159 | 185 | 211 |
| 2018 | 26 | 43 | 56 | 74 | 88  | 103 | 127 | 151 | 174 | 198 |
| 2019 | 26 | 41 | 54 | 68 | 85  | 102 | 126 | 151 | 174 | 198 |
| 2020 | 26 | 41 | 55 | 69 | 86  | 103 | 127 | 151 | 173 | 196 |
| 2021 | 27 | 40 | 55 | 69 | 86  | 102 | 125 | 148 | 174 | 200 |
| 2022 | 26 | 39 | 52 | 67 | 84  | 100 | 121 | 142 | 171 | 200 |
| 2023 | 26 | 39 | 53 | 67 | 81  | 99  | 122 | 145 | 175 | 205 |

Table 2: NHC Cone of Uncertainty values in nautical miles for the Eastern/Central Pacific basin for each year and lead time used in this study. For leadtimes 84 and 108, the earlier and later lead times are averaged. Rows are year, columns are lead times (in hours).

|      | 12 | 24 | 36 | 48 | 60 | 72  | 84  | 96  | 108 | 120 |
|------|----|----|----|----|----|-----|-----|-----|-----|-----|
| 2013 | 30 | 49 | 66 | 82 | 96 | 111 | 134 | 157 | 177 | 197 |
| 2014 | 30 | 46 | 62 | 79 | 92 | 105 | 129 | 154 | 172 | 190 |
| 2015 | 26 | 42 | 54 | 69 | 84 | 100 | 121 | 143 | 162 | 182 |
| 2016 | 27 | 42 | 55 | 70 | 85 | 100 | 118 | 137 | 154 | 172 |
| 2017 | 25 | 40 | 51 | 66 | 79 | 93  | 104 | 116 | 133 | 151 |
| 2018 | 25 | 39 | 50 | 66 | 80 | 94  | 109 | 125 | 143 | 162 |
| 2019 | 25 | 38 | 48 | 62 | 75 | 88  | 101 | 115 | 130 | 145 |
| 2020 | 25 | 38 | 51 | 65 | 78 | 91  | 103 | 115 | 126 | 138 |
| 2021 | 25 | 37 | 51 | 64 | 77 | 89  | 101 | 114 | 126 | 138 |
| 2022 | 25 | 38 | 51 | 65 | 79 | 93  | 106 | 120 | 133 | 146 |
| 2023 | 25 | 38 | 51 | 63 | 78 | 86  | 98  | 110 | 123 | 137 |

## S6 Atlantic Landfall CRPS Comparison

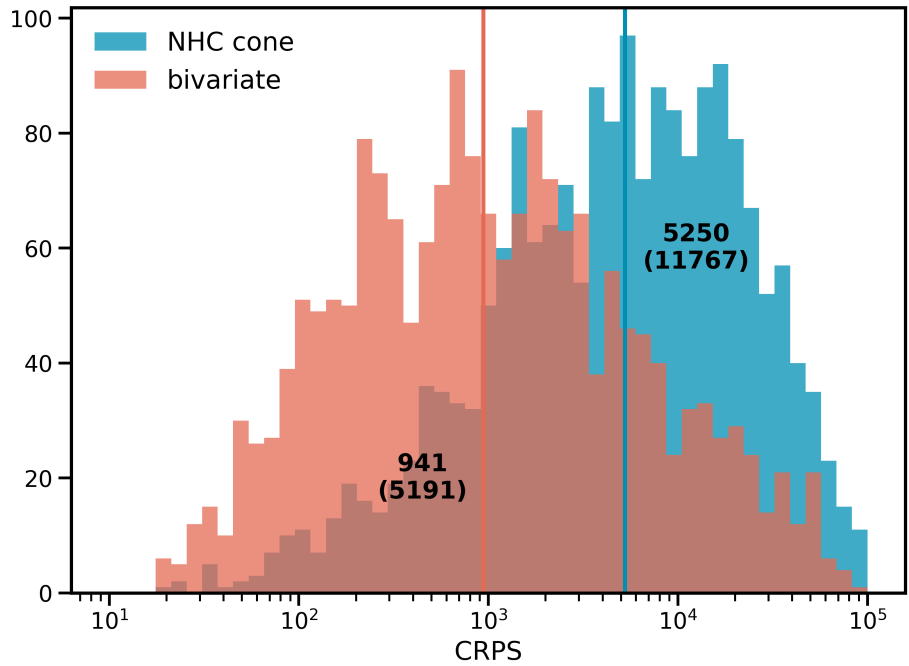

Figure S17: CRPS distribution for bivariate predictions and the NHC cone of uncertainty. Bolded values are the median (mean) of the distributions (left values for bivariate, right values for the NHC cone). A lower CRPS value is better.



\end{document}